\documentclass[12pt, centerh1]{article}
  \usepackage{metalogo} %for XeLaTeX logo
  \usepackage{algpseudocodex} % for the algorithm in Algorithm~1

  \usepackage{color}
  \usepackage{array}
  \usepackage{soul}
  \usepackage{enumitem} 
  \usepackage{bm}
  \usepackage{amsmath}
  \usepackage{natbib}
  \usepackage{hyperref}
  \usepackage{amssymb}
  \usepackage{amsthm}

 \usepackage{graphicx}
  \graphicspath{%
    {converted_graphics/}% inserted by PCTeX
    {Figures-2016-06-13/}% inserted by PCTeX
}

\newtheorem{theorem}{Theorem}[section]

\newtheorem{proposition}[theorem]{Proposition}

\newtheorem{definition}[theorem]{Definition}

\definecolor{jade}{rgb}{0.0, 0.66, 0.42}

\newcommand{\E}{{\rm E}}
\newcommand{\Prob}{{\rm P}}
\newcommand{\Var}{{\rm var}}
\newcommand{\Cov}{{\rm cov}}
\newcommand{\corr}{{\rm cor}}

  \title{A new multivariate Poisson model}
  \author{Orla A. Murphy\textsuperscript{1} and Juliana Schulz\textsuperscript{2}*}
  \date{\small \textsuperscript{1}Department of Mathematics and Statistics, Dalhousie University, Halifax, Canada\\
  \textsuperscript{2}Department of Decision Sciences, HEC Montr\'eal, Montr\'eal, Canada\\
  *Corresponding author. E-mail: juliana.schulz@hec.ca}

\begin{document}
\maketitle
\begin{abstract}
%Insert your abstract here; it should be up to twenty lines long in this format. Avoid symbols as much as possible. Formulas are strongly discouraged, and citations should be avoided. 
%The title and the abstract should be concise and descriptive; the title should not exceed two lines.

Multi-dimensional data frequently occur in many different fields, including risk management, insurance, biology, environmental sciences, and many more. In analyzing multivariate data, it is imperative that the underlying modelling assumptions adequately reflect both the marginal behavior as well as the associations between components. This work focuses specifically on developing a new multivariate Poisson model appropriate for multi-dimensional count data. The proposed formulation is based on convolutions of comonotonic shock vectors with Poisson distributed components and allows for flexibility in capturing different degrees of positive dependence. In this paper, the general model framework will be presented along with various distributional properties. Several estimation techniques will be explored and assessed both through simulations and in a real data application involving extreme rainfall events. \\ \vspace{0.5cm}
%List the key words in alphabetical order.  The MSC 2020 subject classification codes can be found here:
%\url{https://mathscinet.ams.org/msnhtml/msc2020.pdf}
%This document outlines guidelines for authors of The Canadian Journal of Statistics. It also explains an example of the use of the \LaTeX\, class \pkg{cjs-rcs-article}.

\end{abstract}
\newpage

%% Body of the article
\section{Introduction}

When modelling multivariate data, both the marginal and joint model specifications must adequately reflect the random behavior of the variables under study. In general, specifying the univariate distribution of each component is a standard statistical task. Indeed, there exists numerous well-known families of distributions that allow to characterize of a wide-range of types of random variables. Defining the joint model is a more challenging task as the multivariate distribution must incorporate both the marginal specifications while adequately capturing the underlying associations amongst components. 

To this end, there are several techniques that can be used for constructing multivariate models which can be adapted to the case of Poisson margins, a review of some of the more standard approaches can be found in \cite{Inouye_al:2017}. The use of copulas for building multivariate models for discrete data has been explored by several authors, some examples include \cite{VanOphem:1999}, \cite{Pfeifer/Neslehova:2004}, \cite{Nikoloulopoulos/Karlis:2009}, \cite{Smith/Khaled:2012}, \cite{Panagiotelis/Czado/Joe:2012}. Note that when dealing with discrete margins, the use of copulas is more complex, as detailed in \cite{Genest/Neslehova:2007}. Another approach for generating correlated Poisson random variables is to consider mixed models, see, for example, \cite{Aitchison/Ho:1989}, \cite{Karlis/Xekalaki:2005}, \cite{Sarabia:2011}, \cite{Gomez-Deniz:2012}, \cite{Sellers:2016}, and \cite{Karlis/Meligkotsidou:2007}, where the latter considers a finite mixture model.  

A more common approach for modelling correlated count data is to consider a multivariate reduction technique wherein common shock variables are used to induce dependence. In the bivariate setting, this reduces to defining a pair of correlated Poisson random variables as $(X_1,X_2)=(Y_1+Z,Y_2+Z)$, where $Y_1$, $Y_2$, $Z$ are independent Poisson random variables, see, e.g., \cite{Campbell} and \cite{Holgate:1964}. In this construction, the rate of the latent Poisson random variable $Z$, referred to as the common shock variable, regulates the strength of the association between the random pair. A straightforward multivariate extension consists of setting $(X_1,\ldots,X_d)=(Y_1+Z,\ldots,Y_d+Z)$ where $Y_1,\ldots,Y_d,Z$ are independent Poisson random variables, see, e.g., \cite{Loukas/Papageorgiou:1991}, \cite{Loukas:1993}, \cite{Tsionas:1999}, \cite{Karlis:2003}. As in the bivariate case, in the $d$ dimensional setting, the common shock variable $Z$ regulates the strength of the dependence in the model.

A more flexible generalization of this approach is to consider several common shock variables in the construction. For example, the trivariate setting, explored by, e.g., \cite{Mahamunulu:1967} and \cite{Kawamura:1976}, sets
\begin{align*}
X_1 &= Y_1+Y_{12}+Y_{13}+Y_{123} \\
X_2 &= Y_2+Y_{12}+Y_{23}+Y_{123} \\
X_3 &= Y_3+Y_{13}+Y_{23}+Y_{123}
\end{align*}
where each component of $\mathbf{Y}=(Y_1,Y_2,Y_3,Y_{12},Y_{13},Y_{23},Y_{123})$ is an independent Poisson random variable. A general multivariate reduction technique is considered by \cite{Karlis/Meligkotsidou:2005}, who extend the framework to arbitrary dimension $d$ by setting $\mathbf{X} = \mathbf{A} \mathbf{Y}$, where $\mathbf{Y}$ is a $\ell$-dimensional vector with independent Poisson distributed components and $\mathbf{A}$ is a $d \times \ell$ matrix where each element $A_{jk} \in \{0,1\}$, $j=1,\ldots,d$, $k=1,\ldots,\ell$. In this setting, the random vector $\mathbf{Y}$ consists of the set of common shock variables used to generate correlated Poisson random vectors $\mathbf{X}$. Note that a similar formulation can also be considered for constructing a multivariate negative binomial model, see, e.g., \cite{Shi:2014}.

These variants of multivariate Poisson models based on common shocks are commonly used as the formulation is both intuitive and interpretable. An important drawback, however, of the multivariate reduction technique based on common shock variables is that the resulting correlation structure is constrained to fall within a limited range. It is clear that only positive dependence can be characterized in this formulation and, furthermore, it does not allow for strong degrees of dependence, see, e.g., \cite{Genest/Mesfioui/Schulz:2018} and \cite{Schulz/Genest/Mesfioui:2021}. 

To address this limitation, \cite{Genest/Mesfioui/Schulz:2018} proposed a bivariate Poisson model based on comonotonic and counter-monotonic shocks. Together, these two classes of bivariate Poisson models allow for full flexibility in terms of the underlying dependence structure. Indeed, the comonotonic and counter-monotonic shock formulations lead to pairwise correlations spanning the entire spectrum of possible degrees of association, ranging from perfect negative dependence (counter-monotonicity) to perfect positive dependence (comonotonicity). While the comonotonic shock bivariate Poisson model was extended to higher dimensions in the work of \cite{Schulz/Genest/Mesfioui:2021}, as the authors point out, the proposed formulation implies certain limitations on the underlying correlation structure.

The goal of this manuscript is to revisit the idea of using comonotonic shock variables as the building blocks in defining a multivariate Poisson model. In particular, the present paper considers a formulation akin to the multivariate reduction technique, but rather than using independent shock variables, the proposed formulation will incorporate multiple comonotonic shock vectors. As will be demonstrated, the proposed model allows for flexibility in constructing positively correlated Poisson random variables. Moreover, the proposed construction allows for an intuitive interpretation of the source and impact of dependence in the model.

This paper is organized as follows: First, details on the model formulation and resulting properties will be given in \autoref{Sec:model}. Parameter estimation is discussed in \autoref{Sec:estimation}, where several different techniques are presented. \autoref{Sec:simulations} then explores the proposed model through several simulation studies, while \autoref{Sec:data} illustrates the use of the model in a real data application involving extreme rainfall events. Concluding remarks are then provided in \autoref{Sec:conclusion}.

\section{The model}\label{Sec:model}

The proposed multivariate Poisson model is based on convolutions of vectors of comonotonic shocks. Before formally describing the proposed formulation, we begin with some definitions and notation. 

A $d$-dimensional random vector $(Z_1,\ldots,Z_d)$ is said to be comonotonic if and only if each element can be expressed in terms of a common underlying standard uniform random variable, see, e.g., \cite{Dhaene_al:2002}. That is, for a standard uniform random variable $U \sim \mathcal{U}(0,1)$, each $Z_i = F_i^{-1}(U)$, where $F_i$ denotes the marginal distribution function of component $Z_i$, $i=1,\ldots,d$. It can be shown that the comonotonic random vector $(Z_1,\ldots,Z_d)$ has corresponding distribution given by:
\begin{equation}
\label{frechet_upper}
\Prob(Z_1 \leq z_1,\ldots,Z_d \leq z_d) = \min \left\{F_1(z_1),\ldots,F_d(z_d) \right\}
\end{equation}
for any $(z_1,\ldots,z_d) \in \mathbb{R}^d$. We will denote this by $\mathbf{Z} \sim \mathcal{M}_d(F_1,\ldots,F_d)$, where bold font $\mathbf{Z}$ is used to denote a vector. 

Note that the joint distribution given in \eqref{frechet_upper} is referred to as the upper Fréchet-Hoeffding bound. More formally, it can be shown that the joint distribution $H$ of a $d$-dimensional random vector $(X_1,\ldots,X_d)$, with corresponding marginal distributions $F_1,\ldots, F_d$,  satisfies
\begin{equation}
\label{equ:FH_bounds}
\max\{0,\sum_{i=1}^d F_i(x_i)-(d-1)\} \leq H(x_1,\ldots,x_d) \leq \min\{F_1(x_1),\ldots,F_d(x_d)\} .
\end{equation}
The upper Fréchet-Hoeffding bound corresponds to the joint distribution of a random vector exhibiting perfect positive dependence, that is, comonotonicity. In general, the lower bound in Equation~\eqref{equ:FH_bounds} is only a proper distribution in dimension $2$, except under certain specific settings, see, e.g., \cite{Joe:textbook}. A random pair attaining the lower Fréchet-Hoeffding bound is said to exhibit perfect negative dependence, or counter-monotonicity. 

Throughout this paper, we will write $X \sim \mathcal{P}(\lambda)$ for a Poisson-distributed random variable $X$ with mean $\lambda$, and denote the corresponding distribution function, survival function, and probability mass function by $G_{\lambda}$, $\bar{G}_{\lambda}$, and $g_{\lambda}$, respectively.

\subsection{The proposed model}

To build the proposed multivariate Poisson model, we consider $d$ independent latent random vectors $\mathbf{Z}_1,\ldots, \mathbf{Z}_d$, where each random vector $\mathbf{Z}_j$ is itself a $d+1-j$ dimensional comonotonic random vector with Poisson-distributed components.  Thus, while the elements of the random vector $\mathbf{Z}_j$ are perfectly positively dependent, components of distinct vectors, say $\mathbf{Z}_j$ and $\mathbf{Z}_s$, are independent, i.e., $Z_{ij} \perp Z_{rs}$ for any $j \neq s$. We then construct the proposed model in terms of convolutions of these comonotonic shock vectors, as formally presented in the following definition.

\begin{definition}%{Comonotonic shocks multivariate Poisson model}
\label{def:model}
Let $\mathbf{X}=(X_1,\ldots,X_d)$ denote a $d$-variate random vector with Poisson components such that
\begin{align}
\label{equ:model}
\begin{split}
X_1 &= Z_{11}  \\
X_2 &= Z_{21} + Z_{22}  \\
X_3 &= Z_{31} + Z_{32} + Z_{33}  \\
\vdots & \\
X_d &= Z_{d1} + Z_{d2} + Z_{d3} + \cdots + Z_{dd}
\end{split}
\end{align}
where each $Z_{ij}=G_{\omega_{ij}\lambda_i}^{-1}(U_j)$, i.e., $Z_{ij} \sim \mathcal{P}(\omega_{ij}\lambda_i)$, for $j \leq i \in \{1,\ldots,d\}$. The proposed multivariate comonotonic shocks Poisson model will be denoted as $\mathcal{MP}_d(\Lambda,\Omega)$, with $\Lambda=(\lambda_1,\ldots,\lambda_d)$ and $\Omega=(\{\omega_{ij}\}, j \leq i =1,\ldots,d\})$. Note that each $\lambda_j >0$ and $\omega_{ij} \in (0,1)$ for each $j \leq i \in \{1,\ldots,d\}$, subject to the constraint that $\sum_{j=1}^i \omega_{ij}=1$.
\end{definition}

The proposed formulation leads to Poisson-distributed margins $X_i \sim \mathcal{P}(\lambda_i)$ as each $X_i$ consists of a sum of independent $\mathcal{P}(\omega_{ij}\lambda_i)$ random variables, with $\sum_{j=1}^i \omega_{ij}=1$. Thus, the parameters $(\lambda_1,\ldots,\lambda_d)$ represent the marginal Poisson rates, regulating the marginal behaviour of the elements of $\mathbf{X}$. 

The use of the set of comonotonic shock vectors, $\mathbf{Z}_j=(Z_{jj},\ldots,Z_{dj})$, $j=1,\ldots,d$, in the proposed construction induces the dependence structure of $\mathbf{X}$. The parameters $\Omega$ regulate the component-wise weight attributed to each of the $d$ comotononic shock variables, thereby controlling the degree of dependence between the elements of $\mathbf{X}$. In the extremal case where $\omega_{i1}=1 \, \forall j=1,\ldots,d$, $\mathbf{X} = \mathbf{Z}_1$ thereby yielding a comonotonic vector with Poisson margins. On the other hand, when $\omega_{jj}=1 \, \forall j=1,\ldots,d$, the model reduces to that of independence. In between these two extremal cases, varying the values of the weight parameters will accordingly vary the degree of dependence between the components of $\mathbf{X}$. In this sense, the weight parameters can be viewed as dependence parameters in the proposed multivariate model. This will be further explored in \autoref{Sec:depstructure}. Note that the weight parameters $\omega_{ij}$ can also be viewed as the proportion of the marginal mean $\E(X_i)$ due to the $j^{th}$ comonotonic shock, $\mathbf{Z}_j$, since, for each $j \leq i = 1,\ldots,d$ one can write
\[
\omega_{ij} = \E(Z_{ij}) / \E(X_i) =\E(Z_{ij}) / \sum_{j=1}^i \E(Z_{ij}).
\]

\subsubsection{Comparisons with other models}

As discussed in the introduction, the proposed multivariate Poisson model presented in \eqref{equ:model} borrows on the idea of comonotonic shock vectors, as explored in \cite{Genest/Mesfioui/Schulz:2018} and \cite{Schulz/Genest/Mesfioui:2021}. Nonetheless, the proposed formulation presented here is indeed distinct from the latter two works. Indeed, when $d=2$ the proposed model sets 
\[
(X_1,X_2) = \left(G_{\lambda_1}^{-1}(U), G_{\omega_{21}\lambda_2}^{-1}(U) + G_{(1-\omega_{21})\lambda_2}^{-1}(V) \right)
\]
for independent $\mathcal{U}(0,1)$ random variables $U$ and $V$. In contrast, the construction proposed by \cite{Genest/Mesfioui/Schulz:2018} generates pairs of correlated Poisson random variables according to
\[
(X_1,X_2) = \left(G_{\theta \lambda_1}^{-1}(U)+G_{(1-\theta)\lambda_1}^{-1}(V_1), G_{\theta \lambda_2}^{-1}(U)+G_{(1-\theta)\lambda_2}^{-1}(V_2) \right)
\]
where, $U$, $V_1$ and $V_2$ are independent $\mathcal{U}(0,1)$ random variables. Both models build correlated Poisson pairs by considering a form of weighting between independent and comonotonic components, through slightly different representations. Moreover, both models will yield fully flexible (positive) dependence structures ranging from independence (when $\omega_{21}=0$ or $\theta=0$) to comonotonicity (when $\omega_{21}=1$ or $\theta=1$).

In higher dimensions, however, the proposed model allows for further flexibility in contrast to the formulation considered in \cite{Schulz/Genest/Mesfioui:2021}. In particular, \cite{Schulz/Genest/Mesfioui:2021} consider a construction based on a single dependence parameter $\theta$, which represents the weight attributed to a single $d$-variate comonotonic shock vector, while the remaining weight, $1-\theta$, is then attributed to independent Poisson components. This can be represented as
\[
(X_1,\ldots,X_d) = \left( G_{\theta\lambda_1}^{-1}(U)+G_{(1-\theta)\lambda_1}^{-1}(V_1),\ldots, G_{\theta\lambda_d}^{-1}(U)+G_{(1-\theta)\lambda_d}^{-1}(V_d) \right)
\]
for $U,V_1,\ldots,V_d \overset{iid}{\sim} \mathcal{U}(0,1)$. As pointed out by the authors, the fact that a single dependence parameter simultaneously regulates the degree of association amongst all components limits the underlying dependence structure in certain ways. For example, if one pair is independent, then $\theta$ must necessarily be equal to $0$, implying that all components are independent. At the other extreme, if one pair is found to be perfectly positively dependent, then the random vector must be comonotonic as $\theta$ would necessarily be equal to $1$. The proposed model explored in this paper does not suffer from this drawback as the set of weights $\{\omega_{ij}, j<i=1,\ldots,d\}$ allows for further flexibility in terms of the implied pairwise correlations, wherein varying degrees of dependence are attainable amongst different components of the random vector $\mathbf{X}$. This will be further highlighted in \autoref{Sec:depstructure}.

\subsection{Multivariate distribution}

The underlying distribution of the proposed multivariate Poisson model results from the convolutions of the underlying shock vectors, $\mathbf{Z}_1,\ldots,\mathbf{Z}_d$. Recall that for $\mathbf{Z}_j \sim \mathcal{M}_{d+1-j}(G_{\omega_{jj}\lambda_j},\ldots, G_{\omega_{dj}\lambda_d})$, the cumulative distribution is given by
\[
\Prob(Z_{jj}\leq z_{jj},\ldots,Z_{dj}\leq z_{dj}) = \min \left\{ G_{\omega_{jj}\lambda_j}(z_{jj}),\ldots, G_{\omega_{dj}\lambda_d}(z_{dj}) \right\}
\]
for any $(z_{jj},\ldots,z_{dj})\in \mathbb{R}^{d+1-j}$. As shown in, e.g., \cite{Schulz/Genest/Mesfioui:2021}, the corresponding joint probability mass function simplifies to
\begin{equation}
\label{equ:c_fcn}
c_{\Psi_{(j)}}^{(d+1-j)} (z_{jj},\ldots,z_{dj}) = \left[ \underset{k \in \{j,\ldots,d\}}{\min} G_{\omega_{kj}\lambda_k}(z_{kj}) - \underset{k \in \{j,\ldots,d\}}{\max} G_{\omega_{kj}\lambda_k}(z_{kj}-1)  \right]_+ 
\end{equation}
where $[x]_+ = x \mathbf{1}(x>0)$ and $\mathbf{1}(\cdot)$ is the indicator function. In the above, the notation $\Psi_{(j)}$ is used to denote the appropriate subset of the model parameters involved, namely, $\Psi_{(j)}=(\lambda_j,\ldots,\lambda_d,\omega_{jj},\ldots,\omega_{dj})$. By convention, $\Psi = (\Lambda,\Omega)$ will represent the full set of model parameters. The superscript in $c_{\Psi_{(j)}}^{(d+1-j)} $ is used to indicate the dimension of the random vector $\mathbf{Z}_j$.

The joint probability mass function in the proposed multivariate Poisson model, which will be denoted by $f_{\Psi}$, can then be derived by conditioning on the latent comonotonic shock vectors $\mathbf{Z}_1,\ldots,\mathbf{Z}_d$, and making use of \eqref{equ:c_fcn}. In dimension $d$, the probability $\Prob(\mathbf{X}=\mathbf{x})=\Prob(X_1=x_1,\ldots,X_d=x_d)$ is given by
\begin{equation}
\label{equ:f_pmf}
 f_{\Psi}(\mathbf{x}) = \sum_{\mathcal{Z}} \prod_{i=1}^d
c_{\Psi_{(i)}}^{(d+1-i)} (\mathbf{z}_i) 
\end{equation}
where the sum is taken over the set of latent comonotonic shock vectors 
\begin{equation}
\label{equ:set_z}
\mathcal{Z}=\left\{\mathbf{z}_{i}\in \mathbb{N}^{(d+1-i)} : \sum_{j=1}^i z_{ij} \leq x_i,  i=1,\ldots,d \right\},
\end{equation}
with $\mathbb{N} = \{0,1,2,\ldots\}$ representing the set of positive integers, including $0$. Note that in the case where $i=d$, $c_{\Psi_{(d)}}^{(1)} (\mathbf{z}_d) = g_{\omega_{dd} \lambda_d}(z_{dd})$. 

To illustrate, in dimension 2, one obtains
\[
f_{\Psi}(x_1,x_2) = \sum_{z_{22}=0}^{x_2} c_{\Psi_{(1)}}^{(2)} (x_1,x_2-z_{22}) g_{\omega_{22}\lambda_2}(z_{22}),
\]
while in dimension 3 this simplifies to
\begin{multline*}
f_{\Psi}(x_1,x_2,x_3) = \sum_{z_{22}=0}^{x_2}
\sum_{z_{32}=0}^{x_3} \sum_{z_{33}=0}^{x_3-z_{32}} c_{\Psi_{(1)}}^{(3)} (x_1,x_2-z_{22},x_3-z_{32}-z_{33}) \\
\times c_{\Psi_{(2)}}^{(2)} (z_{22},z_{32}) g_{\omega_{33}\lambda_3}(z_{33}).
\end{multline*}

Note that the corresponding cumulative distribution function $F_{\Psi}$ can be derived in a similar manner, by conditioning on the set $\mathcal{Z}$ as in \eqref{equ:set_z}. This leads to 
\[
F_{\Psi}(\mathbf{x}) = \sum_{\mathcal{Z}} 
\left\{\min_{k \in 1, \ldots, d} G_{\omega_{k1}\lambda_k}(z_{k1}) \right\}
\prod_{i=2}^d
c_{\Psi_{(i)}}^{(d+1-i)} (\mathbf{z}_i)  .
\] 
In dimension 3, for example, one obtains 
\begin{multline*}
F_{\Psi}(\mathbf{x}) = \sum_{z_{22}=0}^{x_2}
\sum_{z_{32}=0}^{x_3} \sum_{z_{33}=0}^{x_3-z_{32}}
\min \left\{ G_{\omega_{11}\lambda_1}(x_1),G_{\omega_{21}\lambda_2}(x_2-z_{22}),G_{\omega_{31}\lambda_3}(x_3-z_{32}-z_{33})  \right\} 
\\
\times  c_{\Psi_{(2)}}^{(2)} (z_{22},z_{32}) g_{\omega_{33}\lambda_3}(z_{33}) .
\end{multline*}

\subsubsection{Marginal distributions}

As previously mentioned, it is clear from \eqref{equ:model} that $\mathbf{X} \sim \mathcal{MP}_d(\Lambda,\Omega)$ has Poisson margins, with $X_i \sim \mathcal{P}(\lambda_i)$, $i=1,\ldots,d$. This follows directly from the fact that the Poisson distribution is infinitely divisible. 

The bivariate distribution of any pair $(X_j,X_k)\subset \mathbf{X}$ does not necessarily reduce to that of the $\mathcal{MP}_2$ model, except, of course, for the pair $(X_1,X_2)$. Indeed, consider an arbitrary pair $(X_i,X_j)$ with $i<j \in \{1,\ldots,d\}$. Then, recalling \eqref{equ:model}, we can express the pair in terms of the comonotonic shocks viz.
\begin{align*}
X_i &= Z_{i1} + \cdots + Z_{ii} \\
X_j &= Z_{j1} + \cdots + Z_{ji} + \cdots + Z_{jj}.
\end{align*}
From this, it follows that the joint probability mass function of the components $(X_i,X_j)$, which will be denoted as $f_{i,j}$, can be written as
\begin{equation}
\label{equ:bivariate_pmf}
f_{i,j}(x_i,x_j) = \sum_{\mathcal{Z}_{(i,j)}} \left\{ \prod_{k=1}^i c_{\Psi_{(k)}}^{(2)} (z_{ik},z_{jk})\right\} \times g_{(1-\omega_{j1}-\cdots-\omega_{ji}) \lambda_j}(x_j - \sum_{k \leq i} z_{jk})
\end{equation}
where $\mathcal{Z}_{(i,j)}$ consists of the set of latent comonotonic shocks on components $i$ and $j$, i.e., $\mathcal{Z}_{(i,j)}$ is the set $\{ (z_{i1},\ldots,z_{ii}), (z_{j1},\ldots,z_{ji}): \sum_{r=1}^i z_{ir} \leq x_i, \sum_{s=1}^i z_{js} \leq x_j \}$. Note that the above bivariate probability mass function depends only on a subset of the parameters, which will be denoted as $\Psi_{(i,j)}$. More specifically, for $i<j$, $\Psi_{(i,j)}=(\lambda_k,\omega_{k1},\ldots,\omega_{ki}, k=i,j)$ is the subset of $\Psi$ which is identifiable based on the pair $(X_i,X_j)$. Notice that when marginalizing over the pair $(X_i,X_j)$, for $i<j$, one cannot identify $\omega_{jk}$ for $k>i$. 

This extends to higher dimensional margins as well. That is, there is no common form for the $k$-dimensional marginal distributions. Rather, the distribution of any $k$-variate subset of $\mathbf{X}$ is determined by the specific components included in the subset. 

\subsection{Dependence structure}\label{Sec:depstructure}

As previously discussed, the use of the set of comonotonic shock vectors $\{\mathbf{Z}_j, j=1,\ldots,d\}$ induces the dependence, where the degree of the association is regulated by the corresponding weight parameters $\Omega=(\omega_{ij}, j\leq i \in \{1,\ldots,d\})$. The correlation structure implied by the multiple comonotonic shocks construction will now be explored in more detail. 

\subsubsection{Pairwise correlations}
From \eqref{equ:model}, any arbitrary pair $(X_i,X_j)$ with $i<j \in \{1,\ldots,d\}$ can be expressed as
\[
(X_i,X_j) = \left(\sum_{r=1}^i Z_{ir}, \; \sum_{s=1}^i Z_{js}+\sum_{s=i+1}^{j} Z_{js} \right),
\]
where each $(Z_{ir},Z_{js})$ have respective margins $\mathcal{P}(\omega_{ir}\lambda_i)$ and $\mathcal{P}(\omega_{js}\lambda_j)$, and are comonotonic for $r=s$ and independent for $r \neq s$, $r \in \{1,\ldots,i\}$, $s\in \{1,\ldots,j\}$. Accordingly, only the comonotonic pairs contribute to the pairwise associations. Specifically, the pairwise covariance is given by
\begin{equation}
\label{equ:cov}
\Cov(X_i,X_j) = \sum_{k=1}^{i}  \Cov(Z_{ik},Z_{jk}) 
\end{equation}
and thus the pairwise correlation is given by
%\begin{equation}
%\label{equ:cor}
\[
\corr(X_i,X_j)= \frac{1}{\sqrt{\lambda_i \lambda_j}} \sum_{k=1}^{i}  \Cov(Z_{ik},Z_{jk}) .
\]
%\end{equation}

As shown in \cite{Genest/Mesfioui/Schulz:2018}, the covariance of a comonotonic pair $(Z_{ik},Z_{jk})$, is given by
\begin{equation}
\label{equ:cov_z}
\Cov(Z_{ik},Z_{jk}) = \sum_{m=0}^{\infty} \sum_{n=0}^{\infty} \min\left\{ \bar{G}_{\omega_{ik}\lambda_i}(m), \bar{G}_{\omega_{jk} \lambda_j}(n) \right\} - \omega_{ik}\lambda_i \omega_{jk} \lambda_j.
\end{equation}
Note that the above represents the maximum possible covariance for an arbitrary pair of Poisson-distributed random variables with respective means $\omega_{ik}\lambda_i$ and $\omega_{jk} \lambda_j$, as derived in \cite{Griffiths:1979}.

Let $m_{\lambda_i,\lambda_j}(\omega_{ik},\omega_{jk})$ 
denote the pairwise covariance of the comonotonic pair $(Z_{ik},Z_{jk})$. Following this notation, 
\[
\Cov(X_i,X_j) = \sum_{k=1}^{\min(i,j)} m_{\lambda_i,\lambda_j}(\omega_{ik},\omega_{jk})
\]
For $\mathbf{X} \sim \mathcal{MP}_d(\Lambda,\Omega)$, one then has that
\begin{small}
\[
\Cov(\mathbf{X}) = 
\begin{bmatrix}
\lambda_1 & m_{\lambda_1,\lambda_2}(1,\omega_{21}) & \cdots & m_{\lambda_1,\lambda_d}(1,\omega_{d1}) \\
m_{\lambda_1,\lambda_2}(1,\omega_{21}) & \lambda_2 & \cdots &\sum_{k=1}^2 m_{\lambda_2,\lambda_d}(\omega_{2k},\omega_{dk}) \\
\vdots & \vdots  & \ddots & \vdots \\
 m_{\lambda_1,\lambda_d}(1,\omega_{d1}) & \sum_{k=1}^2 m_{\lambda_2,\lambda_d}(\omega_{2k},\omega_{dk}) & \cdots & \lambda_d \\
\end{bmatrix} .
\]
\end{small}
The corresponding correlation matrix is then given by $\corr(\mathbf{X}) = \mathbf{A} \circ \Cov(\mathbf{X})$, where the matrix $\mathbf{A}$ is the $d \times d$ matrix with $(i,j)^{th}$ element $1/\sqrt{\lambda_i,\lambda_j}$, and the operator $\circ$ is used for the element-wise, or Hadamard, product.

\subsubsection{Strength of association}

The proposed model construction given in \eqref{equ:model} allows for an intuitive interpretation of the strength of the dependence in terms of the weight parameters $\Omega$. As previously discussed, whenever $\omega_{i1}=1\; \forall i=1,\ldots,d$, the components of $\mathbf{X}$ will exhibit perfect positive dependence, that is, $\mathbf{X}$ will be comonotonic. In this case, all elements of $\Cov(\mathbf{X})$ will be maximized in that $\Cov(X_i,X_j)=m_{\lambda_i,\lambda_j}(1,1)$, thereby reaching the largest possible covariance between an arbitrary pair of Poisson random variables with fixed marginal rates $\lambda_i$ and $\lambda_j$, respectively (see \cite{Griffiths:1979}).

Fixing the marginal rates $\Lambda$, each summand in the pairwise covariance given in \eqref{equ:cov}
%i.e., $\Cov(X_i,X_j)=\sum_{k=1}^{\min(i,j)}m_{\lambda_i,\lambda_j}(\omega_{ik},\omega_{jk})$, 
can be shown to be an increasing function of the respective weight parameters $(\omega_{ik},\omega_{jk})$. As the weight parameters are constrained to sum to one component-wise, i.e., for each $i=1,\ldots,d$, $\sum_{j=1}^i \omega_{ij}=1$, it follows that the strength of dependence in the proposed $\mathcal{MP}_d(\Lambda,\Omega)$ model will increase as the weight parameters concentrate along the same comonotonic shock vector, with the extremal case of comonotonicity resulting when $\omega_{i1}=1, \forall i=1,\ldots,d$.

This idea can be formalized by establishing the pairwise positive quadrant dependence (PQD) ordering in the proposed model. In particular, for an arbitrary pair $(X_i,X_j) \subset \mathbf{X} \sim \mathcal{MP}_d(\Lambda,\Omega)$, recall from \eqref{equ:cov} that 
\begin{equation}\label{equ:cov_mfcn}
\Cov(X_i,X_j)=\sum_{k=1}^{\min(i,j)} \Cov(Z_{ik},Z_{jk}) =\sum_{k=1}^{\min(i,j)} m_{\lambda_i, \lambda_j}(\omega_{ik},\omega_{jk})
\end{equation}
We wish to establish that for any $k \in \{1,\ldots
,\min(i,j) \}$ and for fixed $\lambda_i$, $\lambda_j$, $\omega_{ik}$, the function $m_{\lambda_i,\lambda_j}(\omega_{ik},\omega_{jk})$ is increasing in $\omega_{jk}$. While this result is not obvious from the functional form of $m_{\lambda_i,\lambda_j}(\omega_{ik},\omega_{jk})$, as given in \eqref{equ:cov_z}, it follows immediately from the pairwise PQD ordering. This property can be established from the stochastic representation of the model, as will be formalized in the following Proposition, the proof for which can be found in \autoref{Proof}. For a review of the concept of PQD ordering and the properties that ensue, see, e.g., Chapter~9 of \cite{Shaked/Shanthikumar:2007}.

\begin{proposition}{Pairwise PQD ordering}\label{prop_pqd}
\\
Suppose $(X_i,X_j) \subset \mathbf{X} \sim \mathcal{MP}_d(\Lambda,\Omega)$ and $(X_i^\prime,X_j^\prime)\subset\mathbf{X}^\prime \sim \mathcal{MP}_d(\Lambda,\Omega^\prime)$, where $i<j \in \{1,\ldots,d\}$. Further suppose that $\omega_{i \ell }=\omega_{i \ell }^\prime \; \forall \ell \in \{1,\ldots,i\}$, while $\omega_{j \ell}=\omega_{j \ell}^\prime \; \forall \ell \in \{1,\ldots,i\} \setminus k$. Then $\omega_{jk}<\omega_{jk}^\prime \Rightarrow (X_i,X_j) \prec_{PQD} (X_i^\prime,X_j^\prime)$.
\end{proposition}

As discussed in Chapter~9 of \cite{Shaked/Shanthikumar:2007}, the PQD-ordering also ensures an ordering in terms of measures of dependence. In particular, if $(X_1,X_2) \prec_{PQD} (Y_1,Y_2)$ then $\rho(X_1,X_2) \leq \rho(Y_1,Y_2)$ where $\rho$ here denotes Pearson's correlation. This ordering is also retained under other measures of dependence, such as Kendall's $\tau$ and Spearman's $\rho$. 

In the proposed multivariate Poisson model, \autoref{prop_pqd} then implies that for any $0 \leq \omega_{jk} < \omega_{jk}^{\prime} \leq 1$, we have that $\Cov(X_i,X_j) \leq \Cov(X_i^\prime,X_j^\prime)$. More specifically, for fixed $\lambda_i$, $\lambda_j$, $\omega_{ik}$, the function $m_{\lambda_i,\lambda_j}(\omega_{ik},\omega_{jk})$ is increasing in $\omega_{jk}$. In particular, when $\omega_{jk}=0$, the covariance is null, and as $\omega_{jk}$ approaches $1$, the covariance attains its maximum value, given that the remaining parameters are held fixed. This result is particularly helpful for carrying out estimation via the method of moments, as will be detailed in \autoref{Sec:MM}.

\section{Estimation}\label{Sec:estimation}

Suppose a random sample $\mathbf{X}_1,\ldots,\mathbf{X}_n$ is observed, where each $k=1,\ldots,d$, $\mathbf{X}_k = (X_{k1},\ldots,X_{kd}) \sim \mathcal{MP}_d(\Psi)$. It is then of interest to estimate the model parameters, $\Psi = \{ \lambda_i, \omega_{ij}, j<i=1,\ldots,d \}$, with each $\lambda_i>0$ and $\omega_{ij} \in (0,1)$, subject to the constraint $\sum_{j=1}^i \omega_{ij}=1$. There are various methods that lead to consistent estimators for $\Psi$, as will now be explored. In particular, both moment-based and likelihood-based methods will be considered.

The method of moments (MM) approach considered here incorporates marginal first moments and pairwise mixed moments, and allows for a sequential implementation thereby simplifying the computational complexity. Several likelihood-based techniques will also be explored, specifically maximum likelihood (ML) estimation, a two-step ML technique (2S), and a sequential ML approach (SQ). The two-step ML method separates estimation of the marginal parameters, $\Lambda$, from that of the dependence parameters, $\Omega$, into two successive optimization problems. The sequential ML approach further simplifies the estimation procedure by carrying out a series of sequential univariate optimizations for each component of $\Psi$. 

In what follows, detailed descriptions of these estimation techniques will be provided. Throughout, the notation $\mathring{\Psi}$ will be used to denote the MM estimators, $\hat{\Psi}$ for the ML estimators, $\check{\Psi}$ for the two-step estimators, and finally $\tilde{\Psi}$ for the sequential-likelihood estimators. 

\subsection{Method of moments}\label{Sec:MM}

Recall that for $\mathbf{X} \sim \mathcal{MP}_d(\Psi)$, each component is Poisson-distributed with respective marginal rates $\lambda_1,\ldots,\lambda_d$. As such, a consistent estimator for $\Lambda$ is given by $(\mathring{\lambda}_1,\ldots,\mathring{\lambda}_d) = (\bar{X}_1,\ldots,\bar{X}_d)$. 

As previously explored, the weights $\Omega$ are dependence parameters in the sense that they regulate the strength of the correlations between components. Estimation of $\Omega$ can then be carried out by considering mixed sample moments. The pairwise PQD ordering established in \autoref{prop_pqd} ensures that the covariance between components increases as the corresponding pairwise weight parameters increase. More formally, for any $i<j \in \{1,\ldots,d\}$, fixing $\lambda_i>0$, $\lambda_j>0$ and $\omega_{ik} \in (0,1)$, the covariance of the latent component $\Cov(Z_{ik},Z_{jk})=m_{\lambda_i,\lambda_j}(\omega_{ik},\omega_{jk})$ is increasing in $\omega_{jk}$. From this property, a sequential procedure for estimating $\Omega$ can be implemented, as will now be detailed. 

Recall that $\omega_{11}=1$ by definition. In a first step, the weights attributed to the first comonotonic shock vector $\mathbf{Z}_1$ can be established by matching the theoretical pairwise covariance $\Cov(X_1,X_j)=m_{\lambda_1,\lambda_j}(1,\omega_{j1})$ with the sample covariance $S_{1j}=\sum_{m=1}^n (X_{m1}-\bar{X}_1)(X_{mj}-\bar{X}_j)/(n-1)$. 
The MM estimator $\mathring{\omega}_{j1}$ can then be obtained by solving 
\[
S_{1j}=m_{\bar{X}_1,\bar{X}_j}(1,\omega_{j1}) .
\]
The pairwise PQD ordering ensures that this leads to a unique solution whenever $S_{1j} \in [0, m_{\bar{X}_1,\bar{X}_j}(1,1)]$. 

Once the MM estimators $\{\mathring{\omega}_{21},\mathring{\omega}_{31},\ldots,\mathring{\omega}_{d1}\}$ have been established, one can proceed to estimating the weights attributed to the second comonotonic shock vector $\mathbf{Z}_2$ in a similar fashion by considering all pairwise covariances $\Cov(X_2,X_j)$, for $j=3,\ldots,d$. From \eqref{equ:cov_mfcn}, recall that
\[
\Cov(X_2,X_j)=m_{\lambda_2,\lambda_j}(\omega_{21},\omega_{j1})+m_{\lambda_2,\lambda_j}(1-\omega_{21},\omega_{j2})
\]
for $j=3,\ldots,d$. For fixed values of $\lambda_2$, $\lambda_j$, and $\omega_{21}$, the contribution to the covariance due to the second comonotonic shock, $m_{\lambda_2,\lambda_j}(1-\omega_{21},\omega_{j2})$, will be an increasing function of $\omega_{j2}$. Thus, for each $j=3,\ldots,d$, the MM estimator $\mathring{\omega}_{2j}$ is determined as the unique solution to 
\[
S_{2j}-m_{\bar{X}_2,\bar{X}_j}(\mathring{\omega}_{21},\mathring{\omega}_{j1})=m_{\bar{X}_2,\bar{X}_j}(1-\mathring{\omega}_{21},\omega_{j2}).
\]

More generally, at step $k$, the MM estimators of the weights due to the $k^{th}$ comonotonic shock $\mathbf{Z}_k$, i.e. $\{\omega_{jk}, j=k+1,\ldots,d\}$, are established as the unique solution solving
\[
S_{kj}-\sum_{\ell=1}^{k-1} m_{\bar{X}_k,\bar{X}_j}(\mathring{\omega}_{k \ell}, \mathring{\omega}_{j \ell}) = m_{\bar{X}_k,\bar{X}_j} \left(1-\sum_{\ell=1}^{k-1} \mathring{\omega}_{k \ell}, \omega_{j k} \right).
\]

While the implementation of the method of moments allows for a sequential estimation procedure, the estimators $\mathring{\Psi}$ are in fact the simultaneous solutions to
\[
\begin{bmatrix}
\bar{X}_1-\lambda_1 \\
\vdots \\
\bar{X}_d - \lambda_d\\
S_{12}-m_{\lambda_1,\lambda_2}(1,\omega_{21})\\
\vdots \\
S_{d-1 d} - \sum_{k=1}^{d-1} m_{\lambda_{d-1}, \lambda_d}(\omega_{d-1 k},\omega_{d k})
\end{bmatrix}
= \mathbf{0}.
\]
As the above consists of a set of unbiased estimating equations, standard theory ensures that the resulting estimators are consistent and asymptotically normally distributed. This result is further detailed in \autoref{Appendix}, see \autoref{App:MM}, which also provides an expression for the asymptotic variance.

\subsection{Maximum likelihood estimation}\label{Sec:ML}

Recall the joint probability mass function for the $\mathcal{MP}_d(\Lambda,\Omega)$ model, as given in \eqref{equ:f_pmf}. From this, it is straightforward to obtain an expression for the likelihood, $\mathcal{L}(\Psi)$, and log-likelihood, $\ell(\Psi)$, as follows:
\[
\mathcal{L}(\Psi)=\prod_{m=1}^n f_{\Psi}(\mathbf{x}_m), \quad \ell(\Psi) = \sum_{m=1}^n \log f_{\Psi}(\mathbf{x}_m).
\]
The maximum likelihood estimators (MLEs) are then given by
\[
\hat{\Psi} = \underset{\Psi}{\text{arg max}} \; \ell(\Psi),
\]
subject to the constraints that $\lambda_i>0$, $\omega_{ij} \in (0,1)$ and $\sum_{j=1}^i \omega_{ij}=1$ for each $j<i \in \{1,\ldots,d\}$. 

Numerical optimization techniques could be used to solve for $\hat{\Psi}$. Standard likelihood theory ensures that the MLEs are asymptotically $D$-variate Gaussian, more specifically $\sqrt{n}(\hat{\Psi}-\Psi_0) \rightsquigarrow \mathcal{N}_D(\mathbf{0}, \mathcal{I}^{-1})$, where $\Psi_0$ represents the true parameter values, $\mathcal{I}$ is the Fisher information matrix and $D=d+d(d-1)/2$.

\subsection{Two-step likelihood-based estimation}\label{Sec:2S}

Maximum likelihood estimation, as detailed in \autoref{Sec:ML}, involves simultaneously optimizing over a set of $d+{d(d-1)}/{2}$ parameters. Clearly, as $d$ increases, this may become numerically intractable. One approach for simplifying the estimation is to consider a two-step procedure wherein the marginal parameters are first estimated via their respective marginal likelihoods, and the dependence parameters are subsequently estimated based on the joint specification. This approach is particularly convenient in the context of copula models which naturally allows for a factorization of the joint likelihood into the marginal components and an additional term stemming from the copula density. This two-step technique is sometimes referred to as the inference function for the margins (IFM) approach, for further details see, e.g., \cite{Joe:2005, Joe:2014}. It can be shown that under certain regularity conditions, the two-step, or IFM, estimators are consistent and asymptotically Gaussian. This follows directly from the fact that the underlying estimating equations are the score equations stemming from both the marginal and joint model likelihoods. 

More specifically, in the proposed multivariate Poisson model, the two-step estimators are the roots of the set of estimating equations given by
%\begin{equation}\label{equ:IFM_EE}
\[
\mathbf{Q}(\mathbf{X}_1,\ldots,\mathbf{X}_n;\Psi)=\left( \frac{\partial \ell_1(\lambda_1)}{\partial \lambda_1}, \ldots, \frac{\partial \ell_d(\lambda_d)}{\partial \lambda_d}, \frac{\partial \ell(\Lambda,\Omega)}{\partial \Omega^{\top}}  \right)^{\top} = \mathbf{0}
\]
%\end{equation}
where $\ell_i(\lambda_i)$ is the marginal log-likelihood corresponding to component $i$, i.e., $\ell_i(\lambda_i)=\sum_{m=1}^n \log g_{\lambda_i}(x_{mi})$. In implementing this approach, we have a two-step procedure: First, the marginal parameters are estimated via their respective univariate Poisson log-likelihoods, yielding $\check{\lambda}_i=\bar{X}_i$, $i=1,\ldots,d$. In the second step, $\check{\Omega}$ is found as the solution to $\partial \ell(\check{\Lambda},\Omega)/\partial \Omega^{\top}=0$. 

As shown in, e.g., \cite{Joe:2005}, $\check{\Psi}$ is asymptotically Gaussian with $\sqrt{n}(\check{\Psi}-\Psi_0) \rightsquigarrow \mathcal{N}_D \left\{\mathbf{0},V_Q(\Psi_0)\right\}$, where $V_Q(\Psi_0)$ is the Godambe information matrix stemming from the underlying estimating equation $\mathbf{Q}$. . 
Further details are given in \autoref{App:2S} of \autoref{Appendix}.

Note that this two-step estimation procedure leads to a loss in efficiency in comparison to maximum likelihood estimation, see \cite{Joe:2005} for further discussions. Nonetheless, in the proposed model, the two-step method is shown to perform comparably well to ML estimation, as will be detailed in \autoref{Sec:simulations}.

\subsection{Sequential likelihood-based estimation}\label{Sec:SQ}

While the two-step, or IFM, approach detailed in \autoref{Sec:2S} simplifies the numerical complexities of full maximum likelihood estimation, estimation of the dependence parameters $\Omega$ via the joint likelihood still involves a $d(d-1)/2$-dimensional optimization. Further simplifications are possible in re-formulating the underlying estimating equations in terms of marginal and pairwise score equations. The approach considered here can be seen as a special case of a composite likelihood built on marginal and bivariate specifications. The theory of composite likelihood estimation has been discussed by several authors, see e.g., \cite{Varin_al:2011} for an extensive review. \cite{Zhao/Joe:2005}, in particular, discuss the specific setting wherein the composite likelihood is derived from marginal and pairwise specifications, similar to the approach taken here. Note that the sequential implementation that transpires in the $\mathcal{MP}_d(\Psi)$ model will not necessarily hold in general, but is rather a consequence of the specific formulation of the proposed model.

Similarly to the two-step approach detailed in \autoref{Sec:2S}, the marginal parameters can be estimated according to the score equations stemming from the marginal Poisson likelihoods yielding $\tilde{\lambda}_i=\bar{X}_i$, for $i=1,\ldots,d$. The dependence parameters $\Omega$ can then be estimated sequentially based on the pairwise likelihoods of $(X_i,X_j)$, i.e., $\ell(\Psi_{(i,j)};x_i,x_j)=\log f_{i,j}(x_i,x_j)$ stemming from \eqref{equ:bivariate_pmf}, for each $i<j \in \{1,\ldots,d\}$. 

Starting with $\omega_{11}=1$, the first step allows estimation of the weights attributed to the first comonotonic shock $\mathbf{Z}_1$ by considering the bivariate likelihood of $(X_1,X_j)$, for $j=2,\ldots,d$ as follows
\[
\tilde{\omega}_{j1} = \underset{\omega_{j1} \in (0,1)}{\text{arg max }} \ell(\tilde{\lambda}_1,\tilde{\lambda}_j,1,\omega_{j1};\mathbf{x}_1,\mathbf{x}_j).
\]
In the next step, the set of weights associated with the second comonotonic shock vector $\mathbf{Z}_2$ can be determined from the pairwise likelihood of $(X_2,X_j)$, for $j=3,\ldots,d$, as
\[
\tilde{\omega}_{j2} = \underset{\omega_{j2} \in (0,1)}{\text{arg max }} \ell(\tilde{\lambda}_2,\tilde{\lambda}_j,\tilde{\omega}_{21},\tilde{\omega}_{j1},\omega_{j2};\mathbf{x}_2,\mathbf{x}_j).
\]
This procedure can be repeated, until finally reaching $\omega_{dd}$, which is estimated as $\tilde{\omega}_{dd}=1-\sum_{j=1}^{d-1} \tilde{\omega}_{jd}$. Notice that each step in the sequential estimation of both $\Lambda$ and $\Omega$ involves only univariate optimizations, thereby substantially simplifying the numerical complexities. While such an approach results in a loss in efficiency, when compared to maximizing the full model likelihood, simulations (\autoref{Sec:simulations}) suggest that this is rather minor and the proposed sequential likelihood estimators are comparable to the MLEs. 

As was the case with the method of moments, while the estimation procedure described here allows for a sequential implementation, the estimators $\tilde{\Psi}$ are the solutions to the score equations stemming from all marginal and pairwise likelihoods. To see this, let $\mathbf{G}_{0}$ denote the set of $d$ score equations based on the marginal likelihoods, which can be written as
\[
\mathbf{G}_{0}(\mathbf{X};\Lambda) = \left[ \frac{\partial}{\partial \, \Lambda^{\top}} \sum_{j=1}^d \log g_{\lambda_j}(X_j)\right].
\]
Note that $\mathbf{G}_0$ corresponds to the ML score equations when $\Omega=0$, that is, under independence. 

Denote by $\Omega_k$ the set of weights attributed to the $k^{th}$ comonotonic shock $\mathbf{Z}_k$ such that $\Omega_k = \{ \omega_{jk}, j=k+1,\ldots,d\}$. For each $k=1,\ldots,d-1$, define $\mathbf{G}_k(\mathbf{X};\Omega_k)$ as
\[
\mathbf{G}_k(\mathbf{X};\Omega_k) =
\left[
\frac{\partial}{\partial \, \Omega_k^{\top}} \sum_{j=k+1}^d \log f_{k,j}(X_k,X_j)
\right].
\]
That is, $\mathbf{G}_k(\mathbf{X};\Omega_k)$ represents the set of score equations for $\Omega_k$ stemming from the bivariate likelihoods of each pair $(X_k,X_j)$, $j=k+1,\ldots,d$. The sequential, or composite, likelihood estimator $\tilde{\Psi}$ is then the solution to 
\[
\mathbf{G}(\mathbf{X};\Psi)=
\begin{bmatrix}
\mathbf{G}_{0}(\mathbf{X};\Lambda) \\
\mathbf{G}_1(\mathbf{X};\Omega_1) \\
\mathbf{G}_2(\mathbf{X};\Omega_2) \\
\vdots \\
\mathbf{G}_{d-1}(\mathbf{X};\Omega_{d-1}) 
\end{bmatrix}
=\mathbf{0}.
\]

As before, under the usual regularity conditions, standard theory from M-estimation, or estimating equations, ensures that this procedure leads to consistent, asymptotically normal estimators. Further details are provided in \autoref{App:SQ} of \autoref{Appendix}.

\section{Simulations}\label{Sec:simulations}

In order to assess the performance of the estimation techniques described in \autoref{Sec:estimation}, several simulation studies were carried out. Focusing on the trivariate setting, six different scenarios were explored in which varying degrees of associations were considered, including strong, moderate and weak dependence, as induced by the underlying weight parameters $\Omega$. In addition, for each of the three distinct levels of dependence, the mean parameters $\Lambda$ were varied in order to capture the effect of the marginal parameters. Details on the specific scenarios explored are given in what follows. 

\textbf{Scenario 1 - strong dependence: } 
In this setting, the weight parameters were specified as $\omega_{21}=0.9$, $\omega_{31}=0.7$ and $\omega_{32}=0.1$. As such, most of the weight is attributed to the first comonotonic shock, yielding strong levels of dependence across the three components. In scenario 1A, the marginal parameters are set to $\Lambda=(1,2,3)$ thereby yielding 
\[
\corr(\mathbf{X})_{1A} = 
\begin{bmatrix}
 1.00& 0.89 &0.79\\
0.89& 1.00& 0.84\\
0.79 &0.84& 1.00
\end{bmatrix},
\]
while scenario 1B sets $\Lambda=(4,6,8)$ resulting in
\[
\corr(\mathbf{X})_{1B} = 
\begin{bmatrix}
 1.00& 0.93 &0.82\\
0.93 &1.00 &0.87\\
0.82 &0.87& 1.00
\end{bmatrix}.
\]
Notice that although both scenarios are based on the same $\Omega$, $\corr(\mathbf{X})_{1A} \neq \corr(\mathbf{X})_{1B}$ due to the fact that the implied correlation structure depends on the marginal parameters. 

\textbf{Scenario 2 - moderate dependence: }
In the second set of simulations, moderate levels of dependence were explored by setting $\omega_{21}=0.25$, $\omega_{31}=0.1$, $\omega_{32}=0.6$. In this setting, more weight is attributed to the second comonotonic shock so that the second and third components are more strongly correlated. In scenario 2A, setting $\Lambda=(1,2,3)$ leads to
\[
\corr(\mathbf{X})_{2A} = 
\begin{bmatrix} 
1.00& 0.44& 0.28\\
0.44& 1.00 &0.76\\
0.28 &0.76 &1.00
\end{bmatrix}.
\]
Scenario 2B sets $\Lambda=(4,6,8)$ thereby yielding
\[
\corr(\mathbf{X})_{2B} = 
\begin{bmatrix}
 1.00& 0.48& 0.29\\
0.48 &1.00& 0.80\\
0.29 &0.80& 1.00
\end{bmatrix}.
\]

\textbf{Scenario 3 - weak dependence: }
In the final study, weaker levels of dependence were considered wherein less weight was placed on the two comonotonic shocks by setting $\omega_{21}=0.075$, $\omega_{31}=0.075$, $\omega_{32}=0.1$. When $\Lambda=(1,2,3)$ (scenario 3A), the implied correlation structure is given by
\[
\corr(\mathbf{X})_{3A} = 
\begin{bmatrix} 1.00& 0.20& 0.22\\
0.20 &1.00 &0.32\\
0.22 &0.32 &1.00
\end{bmatrix},
\]
while setting $\Lambda=(4,6,8)$ (scenario 3B) results in
\[
\corr(\mathbf{X})_{3B} = 
\begin{bmatrix}
 1.00&  0.24& 0.24\\
0.24& 1.00& 0.34\\
0.24& 0.34 &1.00\end{bmatrix}.
\]

\subsection{Implementation} 

Recall from \autoref{def:model} that $\lambda_j >0$ and $\omega_{ij} \in (0,1)$ for each $j \leq i \in \{1,\ldots,d\}$. Accordingly, the numerical implementations of the four estimation methods must reflect these parameter constraints. In the context of likelihood-based estimation, this can easily be done by considering a reparameterization which removes such constraints. Here, this is accomplished by considering $\eta_j=\log(\lambda_j)$ and $\alpha_{ij}=\log\{\omega_{ij}/(1-\omega_{ij})\}$, for each $j \leq i \in \{1,\ldots,d\}$. Note that the formulation of the model will automatically ensure that $\sum_{j=1}^i \omega_{ij}=1$ for each $i<j \in \{1,\ldots,d\}$, as evaluating a Poisson probability with a negative rate will yield a probability of zero. 

The method of moments requires somewhat more care in ensuring the parameter constraints are respected. It is clear that $\mathring{\lambda}_j=\bar{X}_j>0$ for any $j\in\{1,\ldots,d\}$, and thus the marginal parameter constraints are naturally satisfied. Note that this is also the case in estimating the marginal parameters for the two-step and sequential likelihood-based methods, as $\check{\lambda}_j=\tilde{\lambda}_j=\bar{X}_j$, for $j=1,\ldots,d$. In terms of the weight parameters, more caution is required to ensure that each $\omega_{ij} \in (0,1)$. To see this, first consider the set of weights attributed to $\mathbf{Z}_1$, that is, $\{\omega_{j1}, j=1,\ldots,d\}$. While theoretically $m_{\lambda_1,\lambda_j}(1,\omega_{j1})$ should fall between $0$ and $m_{\lambda_1,\lambda_j}(1,1)$, the sample pairwise covariance could extend beyond these bounds. Thus, in implementing the method of moments, whenever $S_{1j}<0$ the resulting MM estimator will be set to 0, and when $S_{1j}>m_{\bar{X}_1,\bar{X}_j}(1,1)$ the convention will be to set $\mathring{\omega}_{j1}=1$. Similarly, in estimating $\{\omega_{32},\ldots,\omega_{d2}\}$, whenever $S_{2j}-m_{\bar{X}_2,\bar{X}_j}(\mathring{\omega}_{21},\mathring{\omega}_{j1})$ extends beyond the interval $[0,m_{\bar{X}_2,\bar{X}_j}(1-\mathring{\omega}_{21},1)]$ the estimate $\mathring{\omega}_{2j}$ will be restricted to the interval $[0,1]$. More generally, throughout the sequential estimation procedure, the convention will be to bound $\mathring{\omega}_{jk}$ to fall within $(0,1)$, capping the estimate either at $0$ or $1$, as appropriate. Such implementation issues will only be problematic in smaller sample sizes as the sample covariance $S_{ij}$ is a consistent estimator for $\Cov(X_i,X_j)$.

Each of the likelihood-based methods, namely ML, 2S, and SQ, require optimizing an underlying likelihood function. In implementing these optimization procedures, appropriate starting values must be specified. For simplicity, a grid search was considered for the sequential likelihood estimation approach, and the resulting SQ estimates were then used as the starting values for the 2S and ML methods. 

All simulations were carried out using \textsf{R}. In particular, the \texttt{uniroot} function was used for method of moments estimation, while \texttt{optim} was used for all three likelihood-based methods.

\subsection{Simulation results}

\begin{figure}
\begin{center}
\includegraphics[width=0.475\textwidth]{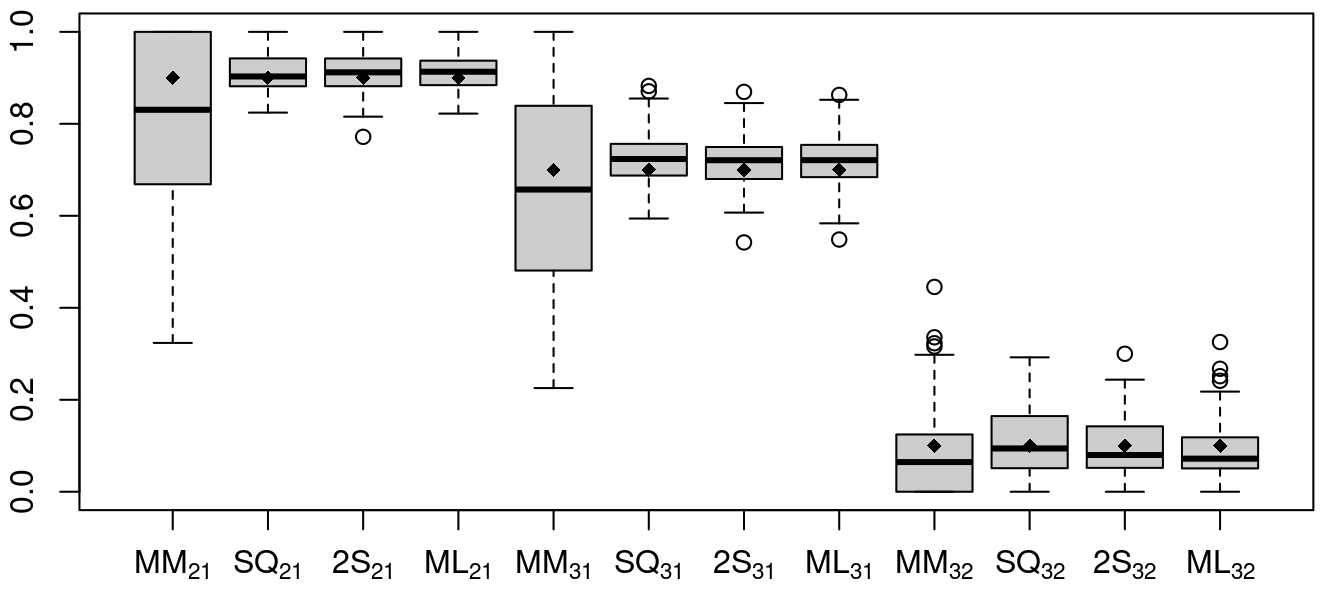}
\includegraphics[width=0.475\textwidth]{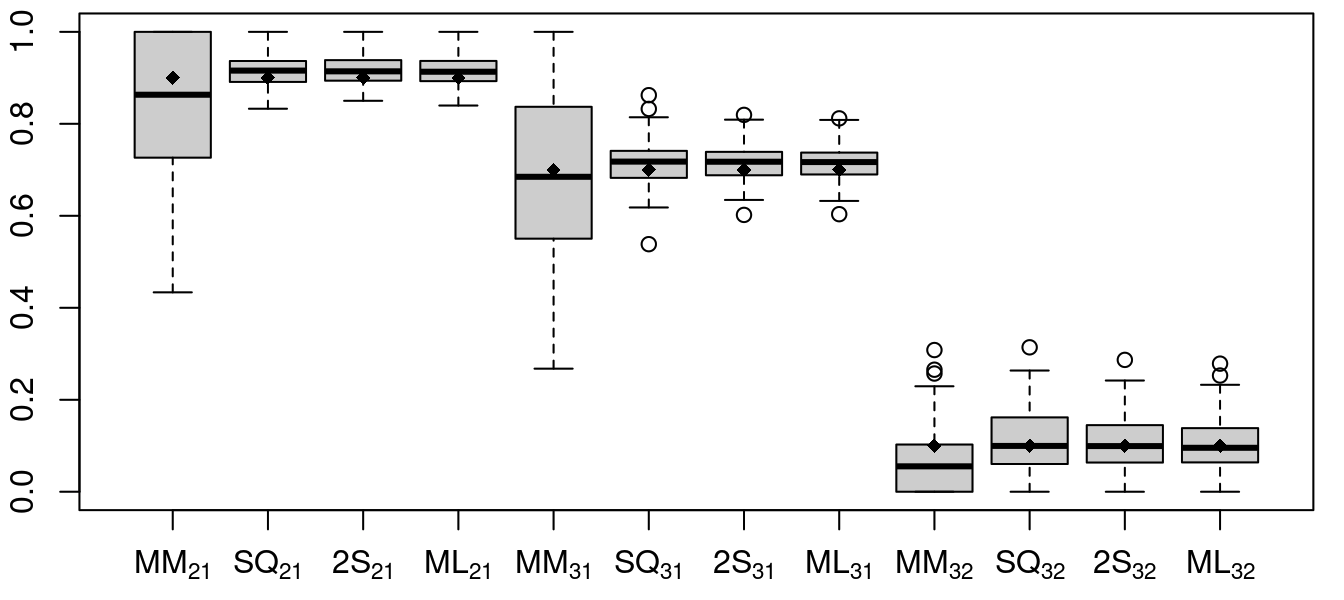}
\includegraphics[width=0.475\textwidth]{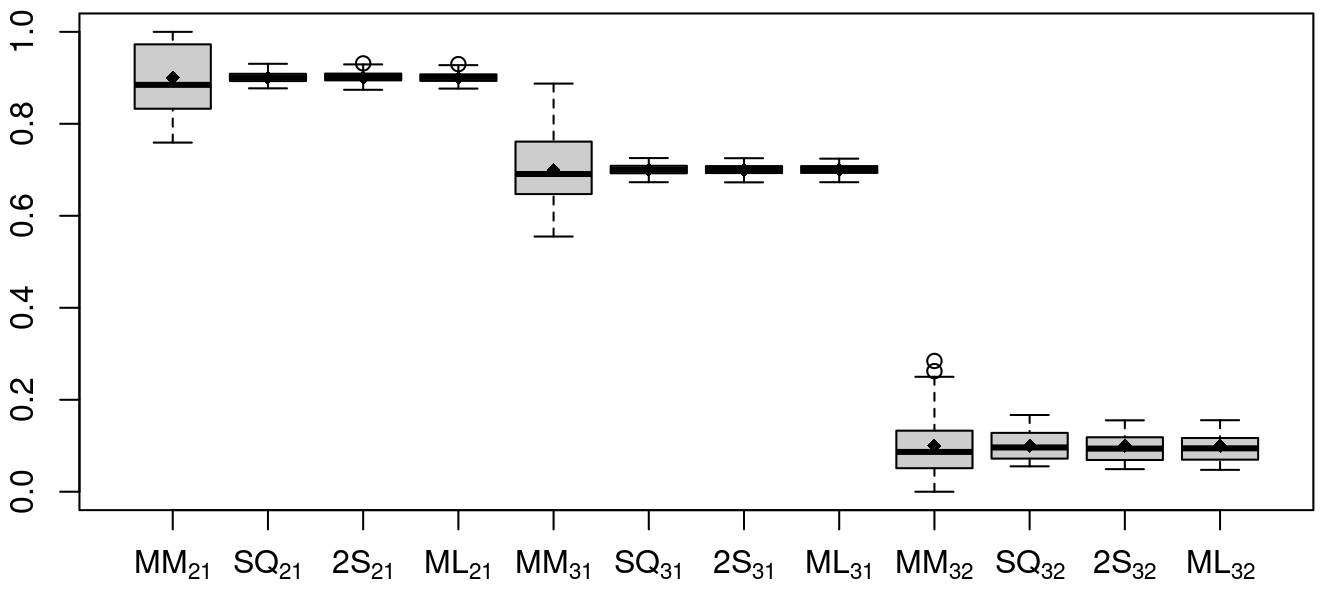}
\caption{Estimates of $\Omega$ for scenario 1A. Top left plot showing results for $n=50$, top right for $n=100$, and bottom for $n=1000$.}
\label{fig:boxplots1A}
\end{center}
\end{figure}

\begin{figure}
\begin{center}
\includegraphics[width=0.475\textwidth]{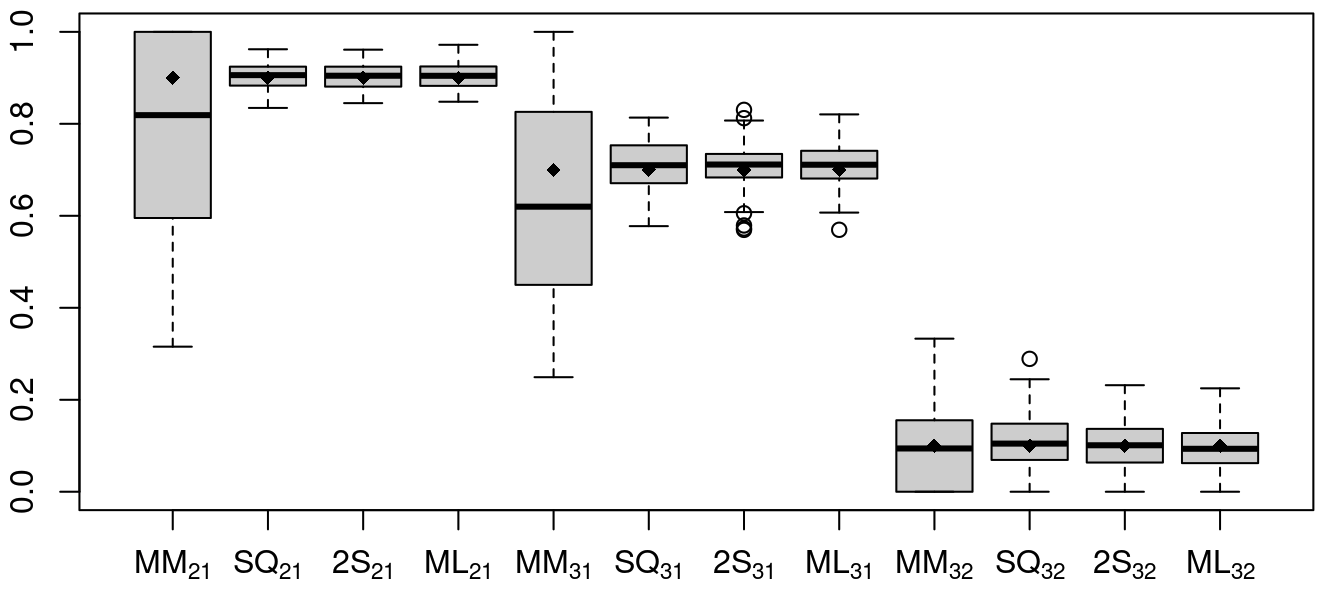}
\includegraphics[width=0.475\textwidth]{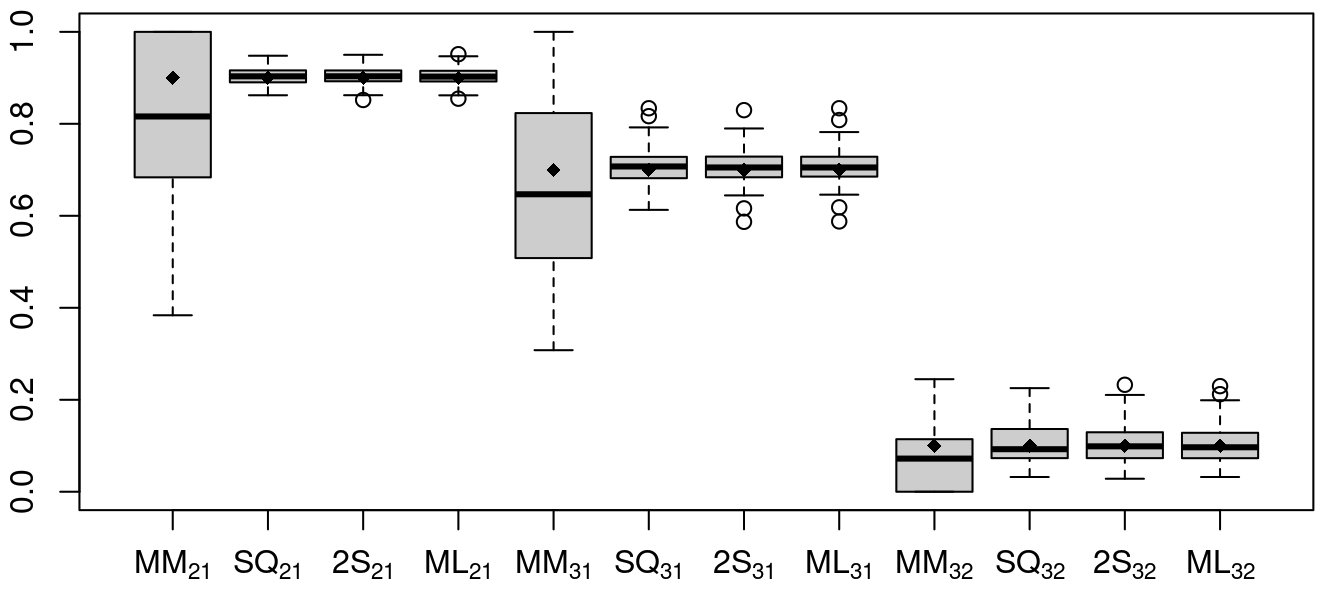}
\includegraphics[width=0.475\textwidth]{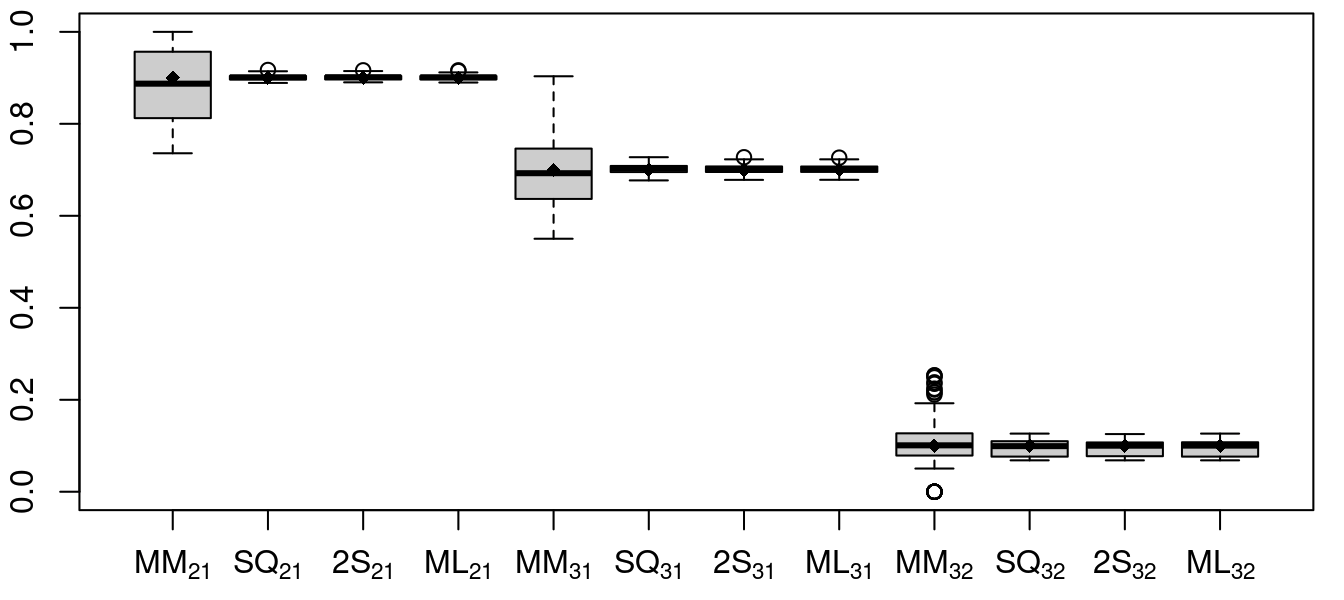}
\caption{Estimates of $\Omega$ for scenario 1B. Top left plot showing results for $n=50$, top right for $n=100$, and bottom for $n=1000$.}
\label{fig:boxplots1B}
\end{center}
\end{figure}

\begin{figure}
\begin{center}
\includegraphics[width=0.475\textwidth]{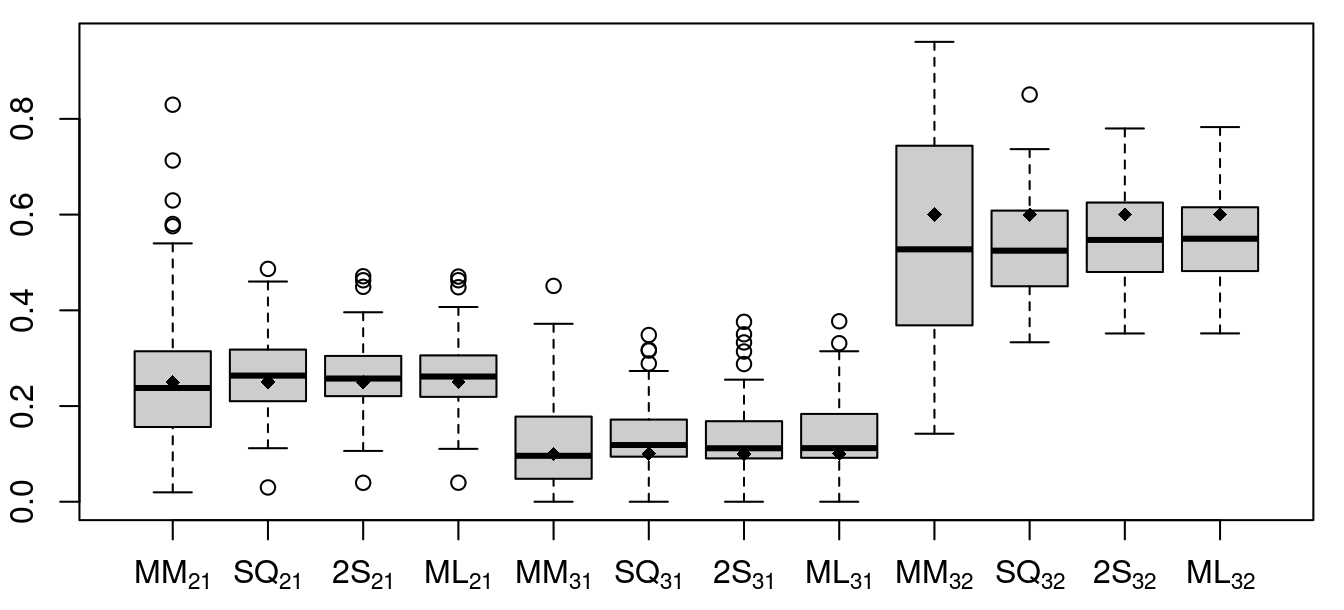}
\includegraphics[width=0.475\textwidth]{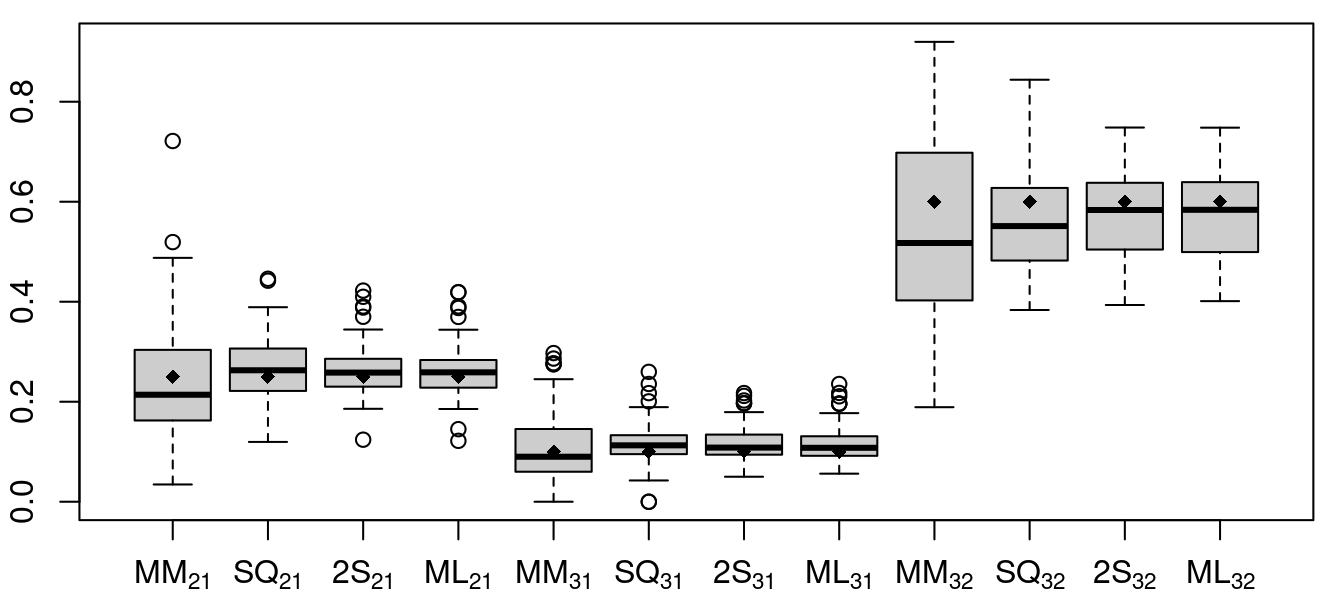}
\includegraphics[width=0.475\textwidth]{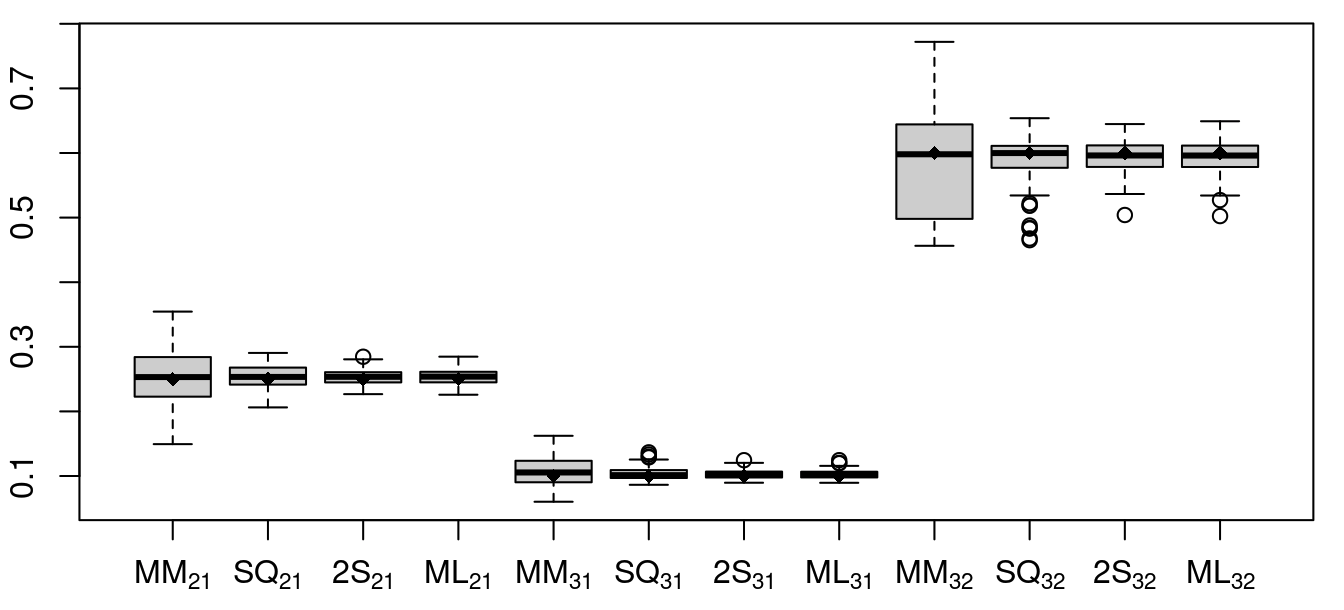}
\caption{Estimates of $\Omega$ for scenario 2A. Top left plot showing results for $n=50$, top right for $n=100$, and bottom for $n=1000$. }
\label{fig:boxplots2A}
\end{center}
\end{figure}

\begin{figure}
\begin{center}
\includegraphics[width=0.475\textwidth]{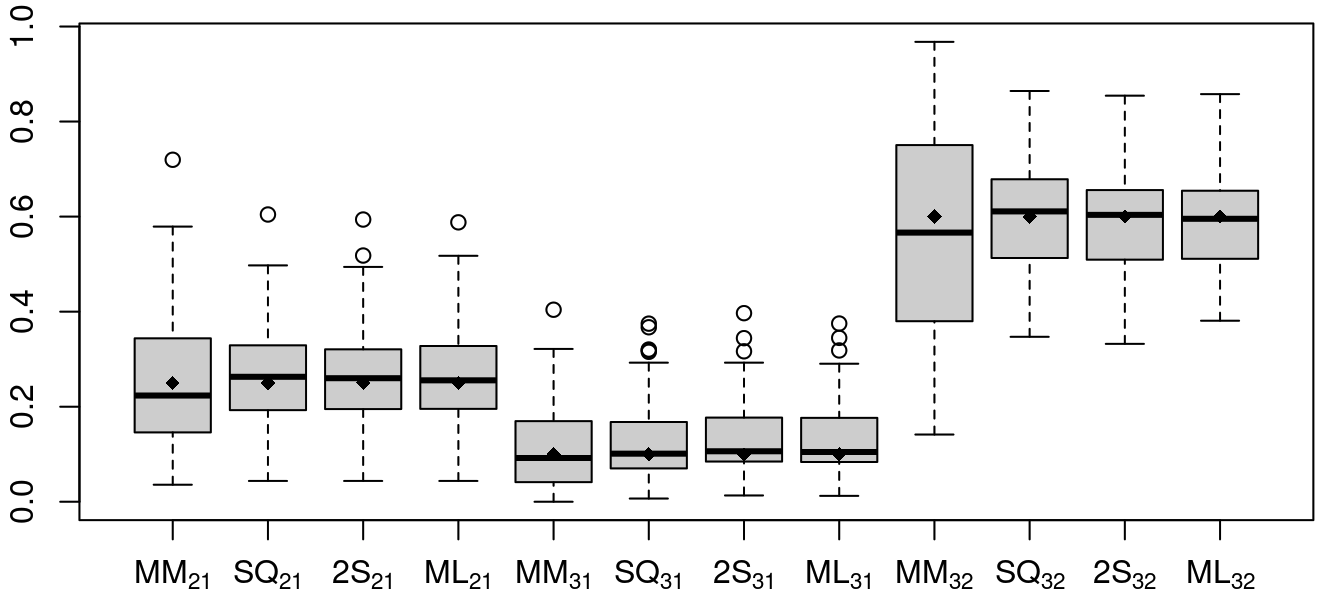}
\includegraphics[width=0.475\textwidth]{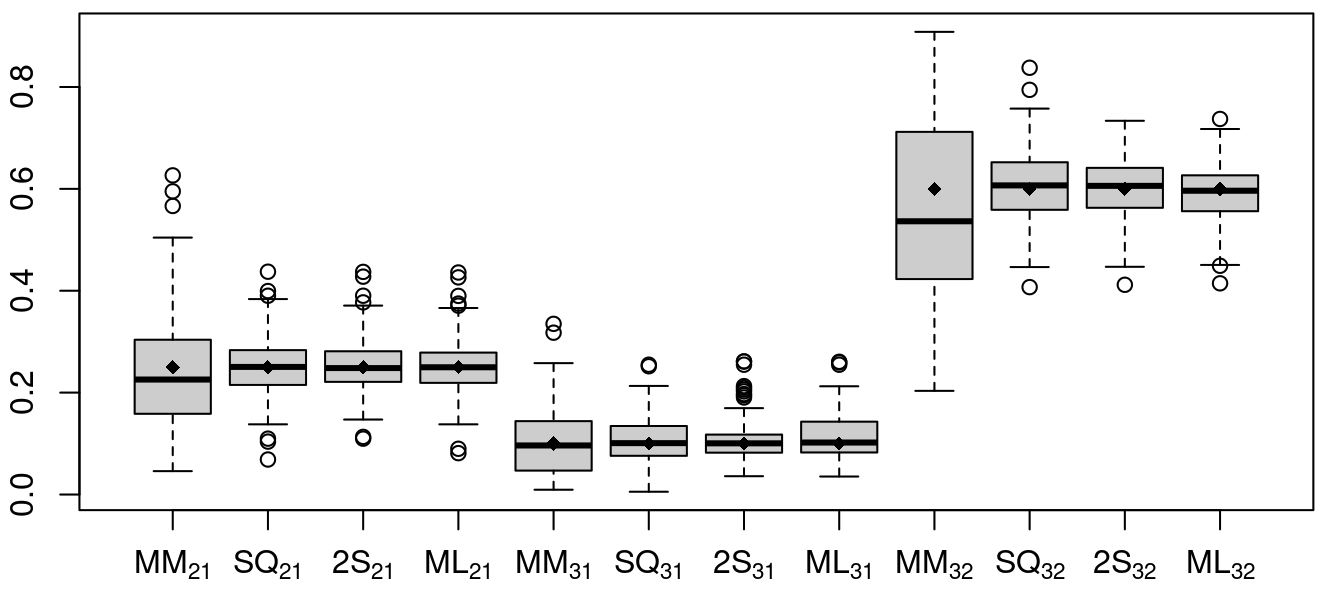}
\includegraphics[width=0.475\textwidth]{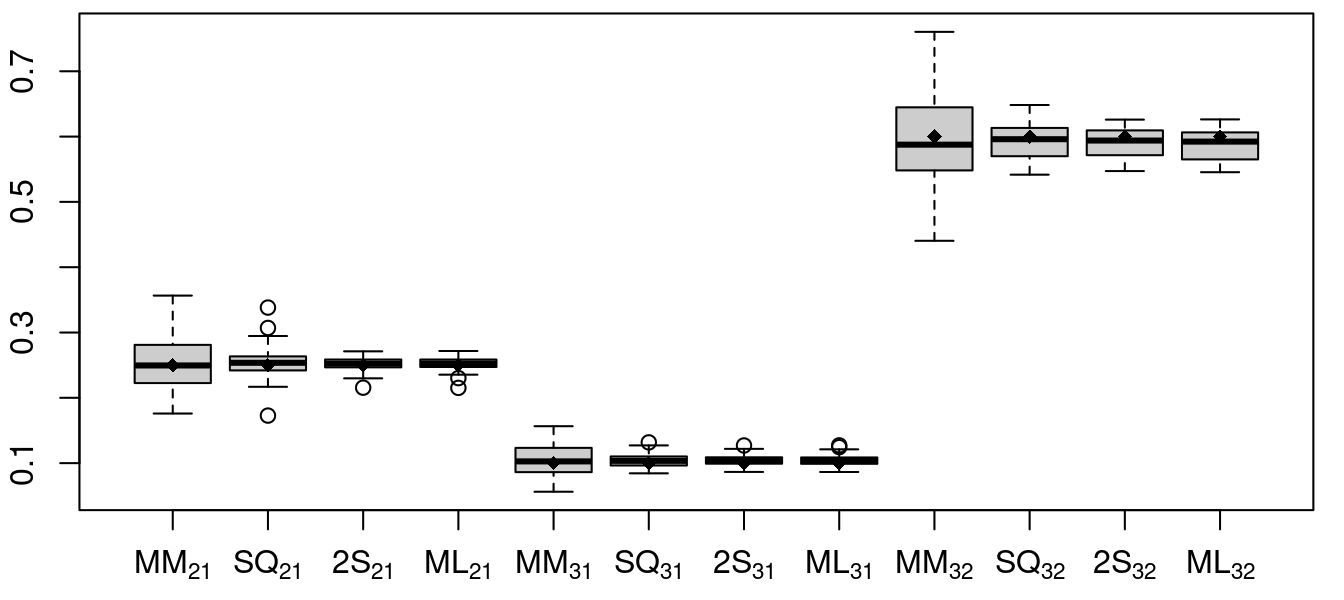}
\caption{Estimates of $\Omega$ for scenario 2B. Top left plot showing results for $n=50$, top right for $n=100$, and bottom for $n=1000$.}
\label{fig:boxplots2B}
\end{center}
\end{figure}

\begin{figure}
\begin{center}
\includegraphics[width=0.475\textwidth]{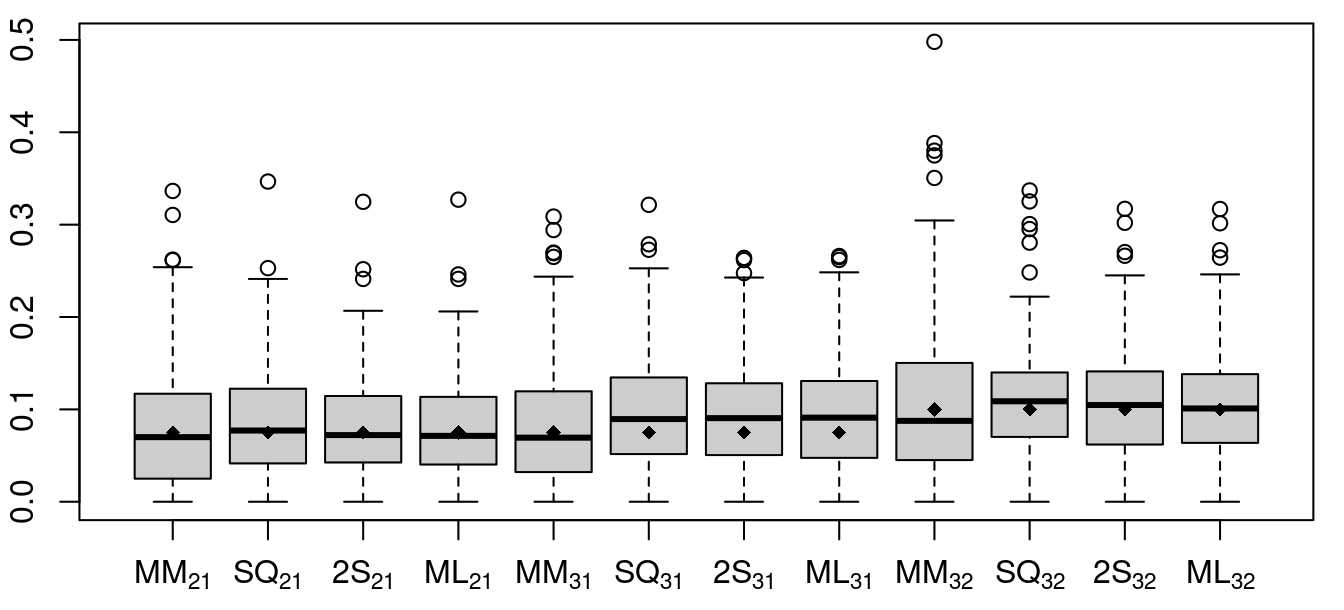}
\includegraphics[width=0.475\textwidth]{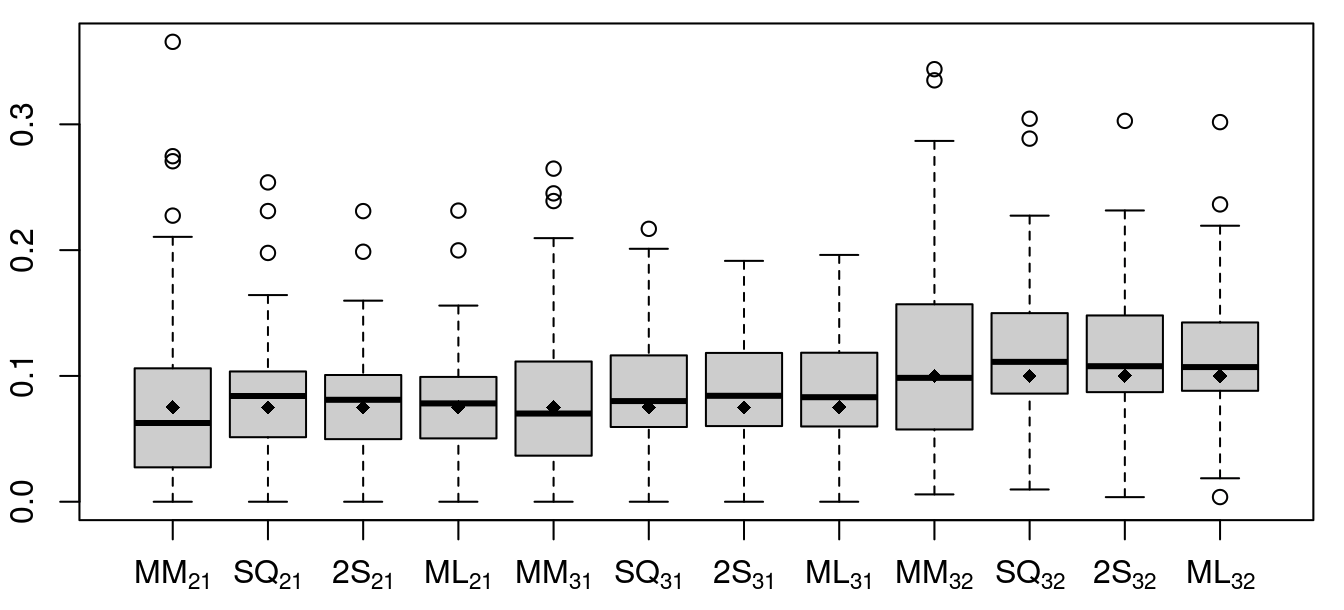}
\includegraphics[width=0.475\textwidth]{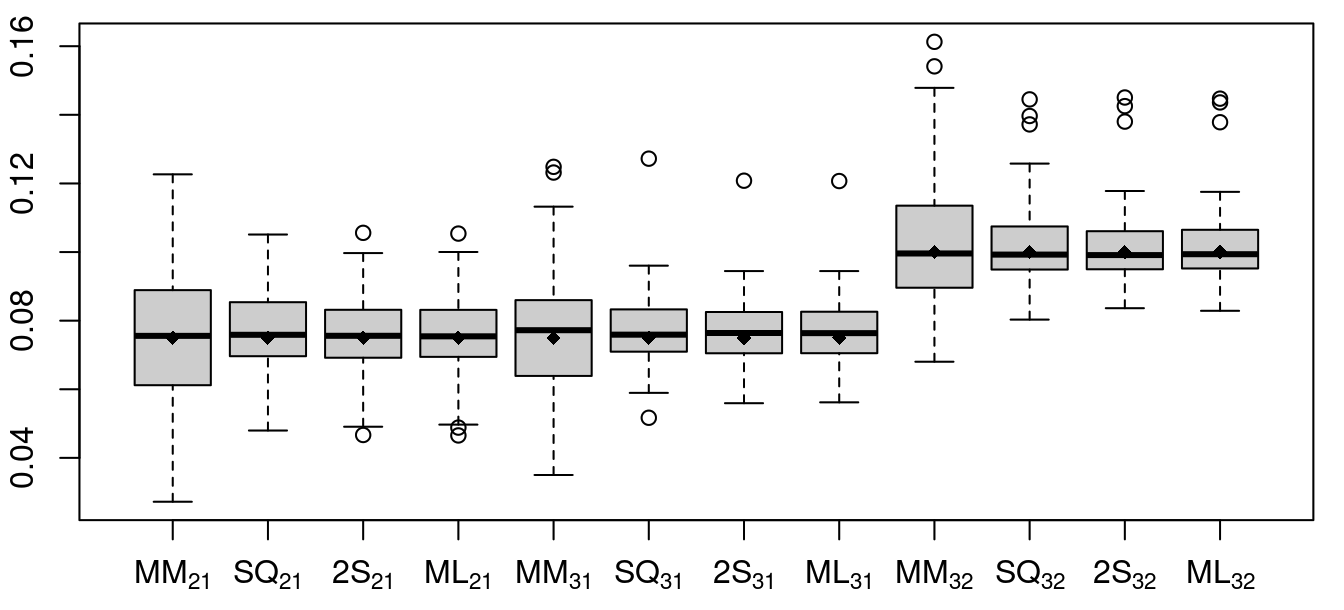}
\caption{Estimates of $\Omega$ for scenario 3A. Top left plot showing results for $n=50$, top right for $n=100$, and bottom for $n=1000$.}
\label{fig:boxplots3A}
\end{center}
\end{figure}

\begin{figure}
\begin{center}
\includegraphics[width=0.475\textwidth]{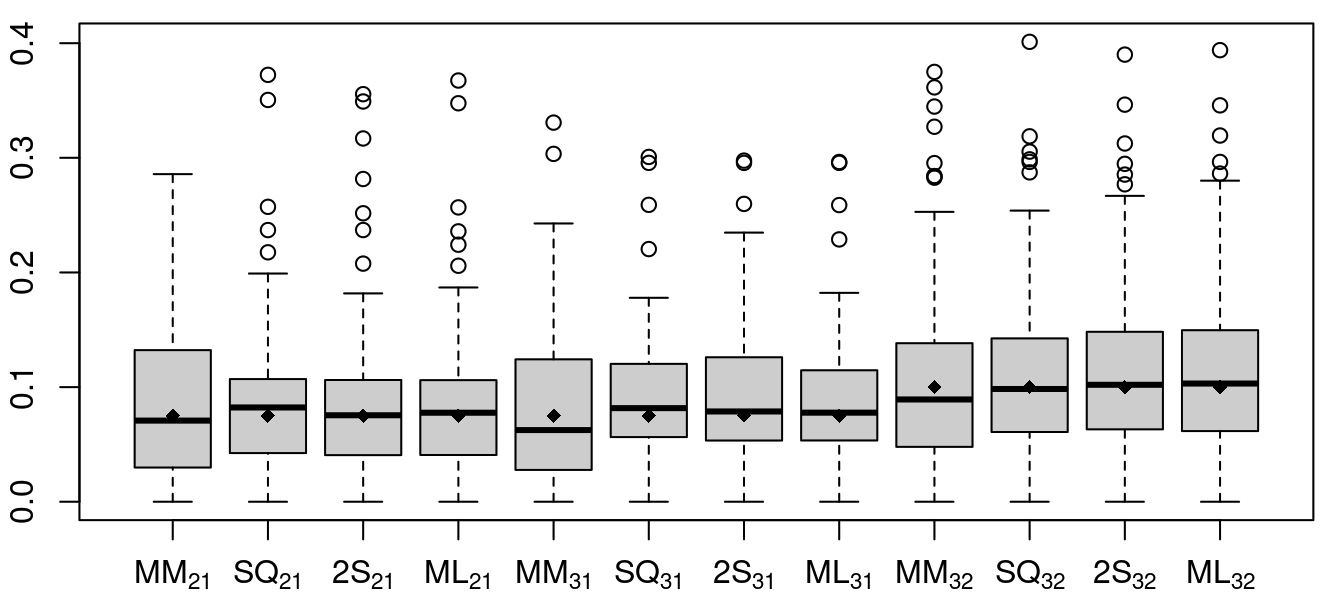}
\includegraphics[width=0.475\textwidth]{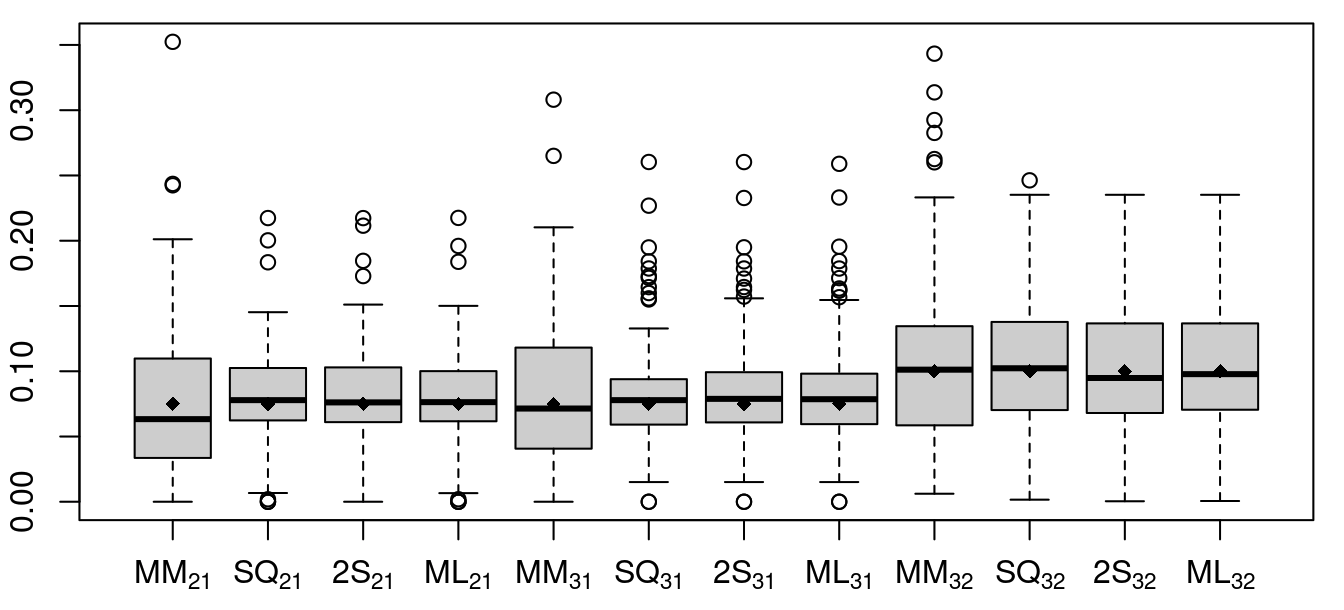}
\includegraphics[width=0.475\textwidth]{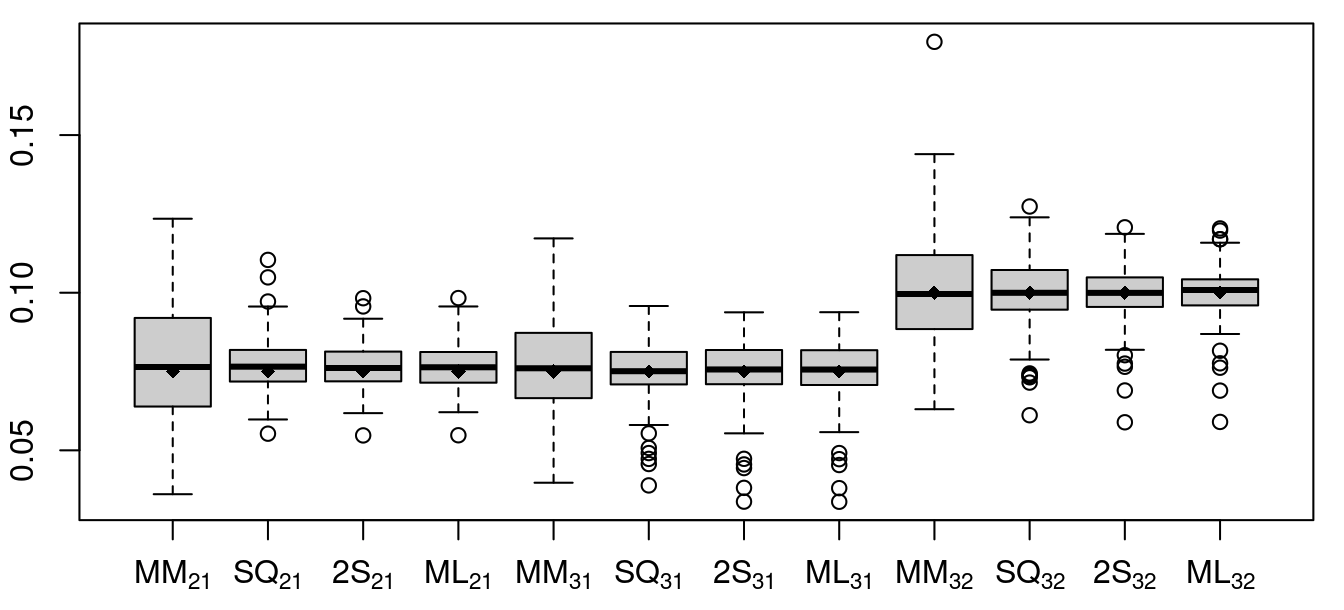}
\caption{Estimates of $\Omega$ for scenario 3B. Top left plot showing results for $n=50$, top right for $n=100$, and bottom for $n=1000$. }
\label{fig:boxplots3B}
\end{center}
\end{figure}

\begin{figure}
\includegraphics[width=0.32\textwidth]{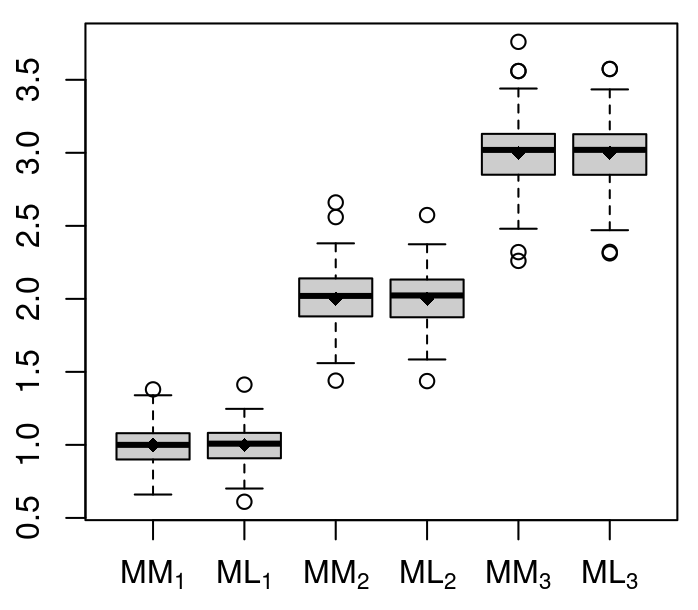}
\includegraphics[width=0.32\textwidth]{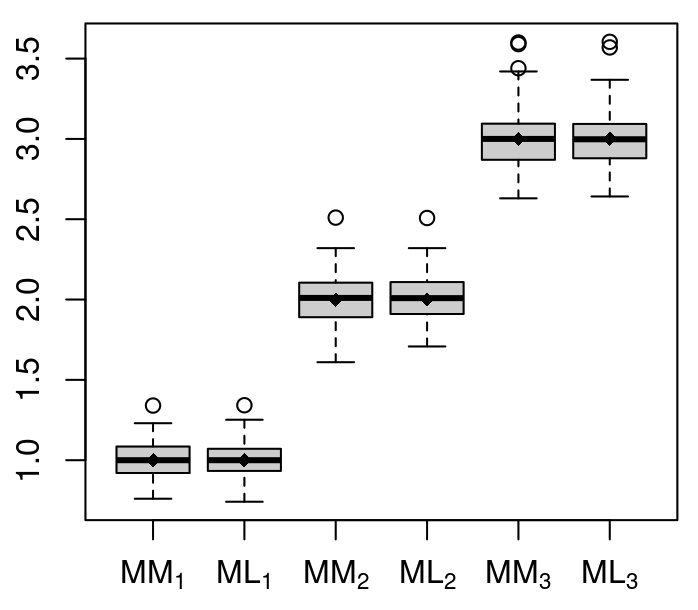}
\includegraphics[width=0.32\textwidth]{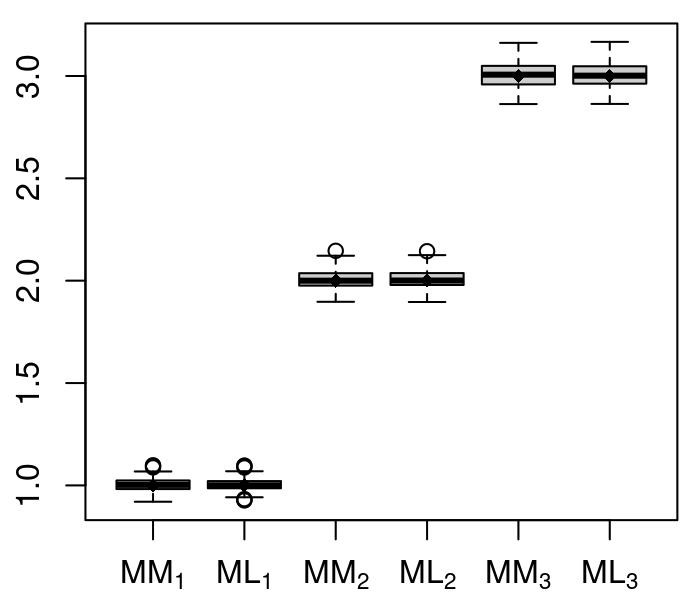}
\caption{Estimates of $\Lambda$ for scenario 1A. Left to right: $n=50$, $n=100$, and $n=1000$. }
\label{fig:boxplotslam}
\end{figure}

Each of the six settings were implemented with sample size varying in $n \in \{50, 100, 1000\}$. Results, summarizing $100$ replications, are provided in \autoref{fig:boxplots1A} through \autoref{fig:boxplotslam}. Throughout, the diamond symbol in the plots is used to indicate the true value of the corresponding parameter. 

\autoref{fig:boxplots1A} through \autoref{fig:boxplots3B} display the estimates of $\Omega$ in scenarios 1A through 3B. It is clear from these results that the method of moments resulted in the most variability, and although it did show significant improvements with increasing sample size, it generally remained more variable than the remaining three estimation approaches across all $n$. It is interesting to remark that the method of moments performed more comparably to the likelihood-based methods for weaker levels of dependence, more specifically, for smaller values of $\omega_{ij}$. This is particularly apparent in scenarios 2 (\autoref{fig:boxplots2A} and \autoref{fig:boxplots2B}) and 3 (\autoref{fig:boxplots3A} and \autoref{fig:boxplots3B}). All three likelihood-based estimation methods (SQ, 2S, and ML) showed similar performance in estimating the dependence parameters. This suggests that the loss in efficiency in both the two-step and sequential procedures is in fact quite minimal. This finding is particularly useful in higher dimensions, wherein complete maximum likelihood estimation may become infeasible. Indeed, as the SQ method involves a series of univariate optimizations, likelihood-based estimation in high dimensional problems could be more easily carried out in considering the sequential approach. Finally, as one would expect, all methods yielded better performance as the sample size increased, with decreasing variability and bias.

As previously mentioned, appropriate starting values must be provided in the optimization procedures used for the likelihood-based estimation methods. While multiple initializations were considered for the sequential approach, the SQ estimates were then used as the starting points for the two-step and full maximum likelihood estimation procedures. It is clearly a more difficult problem to establish suitable starting values in the 2S and ML methods as estimation of all components of $\Omega$ is done simultaneously. It was found that all of the likelihood-based techniques were sensitive to the choice of starting values, with the 2S and ML methods exhibiting increased sensitivity in comparison to the SQ approach. Indeed, using the same initializations as the SQ approach, or even the resulting SQ estimates themselves, could lead to non-convergence for the two-step and full likelihood estimation methods. In the simulations considered here, this occurred in as little as $0$ and up to $16$ of the $100$ replications in a given scenario. 
In general, there were more instances of non-convergence in settings with smaller sample sizes. Note that these non-convergence results are not included in the boxplots. In practice, multiple initializations should be used for the likelihood-based methods. 

Across all scenarios, there is very little difference in the results when the marginal parameters are varied from $\Lambda=(1,2,3)$ (setting A) to $\Lambda=(4,6,8)$ (setting B). Thus, while the marginal specifications certainly affect the implied correlation structure, it seems to have very little impact on the estimation of the dependence parameters $\Omega$. \autoref{fig:boxplotslam} displays the estimates of $\Lambda$ from scenario 1A under the method of moments and full maximum likelihood estimation. Recall that each of the MM, 2S and SQ approaches estimate the marginal parameters $\lambda_j$ by their respective sample means $\bar{X}_j$, for $j=1,\ldots,d$. It is clear from \autoref{fig:boxplotslam} that both approaches yield very similar estimates, with standard errors decreasing with increasing sample size, as expected. 

Finally, the runtimes of the estimation methods were also examined. In order to allow for a more appropriate comparison of the likelihood-based approaches, a common set of initializations were considered for each of the SQ, 2S and ML methods, choosing values close to the true parameter values. It was found that the MM approach was, on average, fastest and remained relatively stable as the sample size increased. In terms of the likelihood-based approaches, the SQ method took, on average, less than half the time required for the two-step technique, which in turn ran, on average, in half the time required to complete full maximum likelihood estimation. It was found that the execution times for the three likelihood-based methods increased with sample size.

\section{Data illustration}\label{Sec:data}

To further examine the usefulness of the proposed model, a real data application was considered. In particular, the proposed multivariate Poisson model was used to model extreme precipitation events in the Maritime region of Canada. The raw data consisted of daily rainfall recordings (in millimeters) from three locations in the Kings and Annapolis Valley counties in Nova Scotia. Both of these counties have suffered from flooding in recent years due to heavy rainfall. The raw data, extending from the years 1919 to 2002, are available from the Environment Canada weather archives. This analysis focused on rainfall data from three weather stations, namely Annapolis Royal (Climate ID 8200100), Springfield (Climate ID 8205200), and Kentville (Climate ID 8202810 and 8202800). (Note that the Kentville weather station location moved in 1996, and as such is characterized by two distinct identification numbers). The raw data were then manipulated in order to handle missing values appropriately. In particular, years which were found to have a high percentage of missing data during the rain season (May-September) at any of the three locations were excluded from the dataset. For years with moderate levels of missingness (around 10\%), nearby stations were used to verify whether any extreme rainfall events had occurred, in which case the year would be removed. Finally, all years with missingness levels under 10\% were retained for the analysis, leading to 71 years of observations. 

The number of yearly exceedances above a high threshold can be shown to follow a Poisson distribution, see e.g., Corollary 4.19 in \cite{resnick2008}. In terms of the precipitation data explored here, thresholds for the rainfall amounts at each of the three stations were selected using stability plots for the parameters of a fitted generalized Pareto distribution, where the latter is used to approximate the distribution of the threshold exceedances. The selected thresholds ranged from the $96^{th}$ to the $98^{th}$ percentile of the daily rainfall amounts. The final dataset then consisted of $71$ trivariate observations of the number of extreme rainfall events wherein the daily rainfall amount exceeded the pre-defined threshold at each of the respective weather stations. Note that observations were declustered in that consecutive days of extreme rainfall were counted as a single extreme event. The marginal Poisson assumption for the resulting number of extreme rainfall occurrences was verified empirically at each of the weather stations using a Chi-squared goodness-of-fit test, yielding p-values greater than $0.3$ for each location. In the analysis here, the counts at the Annapolis Royal station is labelled as $X_1$, Springfield as $X_2$, and Kentville as $X_3$.

All four estimation techniques outlined in \autoref{Sec:estimation} were used to estimate the $\mathcal{MP}_d(\Lambda,\Omega)$ model parameters. \autoref{tab:omega} and \autoref{tab:lambda} display the resulting marginal and dependence parameter estimates, respectively, while \autoref{tab:rho} provides the implied pairwise correlations. \autoref{tab:omega} and \autoref{tab:lambda} also provide bootstrap standard errors (recorded in parentheses), while \autoref{tab:rho} provides $95\%$ bootstrap confidence intervals for the estimated pairwise correlations, based on 1000 bootstrap samples, selected with replacement. 

The results obtained are in line with what was observed in the simulation studies. In terms of the weight parameters $\Omega$, the three likelihood-based methods yield very similar results, both in terms of the estimated values as well as the standard error. The method of moments results in more variability, particularly for estimation of $\omega_{21}$ and $\omega_{31}$. There is seemingly less variability in estimating the third dependence parameter $\omega_{32}$. Note that this phenomena was also observed in the simulations, that is, method of moments estimation performs similar to the likelihood-based approaches when the corresponding weight parameter is small. Estimation of the marginal parameters $\Lambda$, as shown in \autoref{tab:lambda}, is similar in both approaches, namely, when estimation is based on the component-wise behavior (MM, 2S, SQ) and when it is based on the joint behavior (ML), exactly as was observed in the simulations. 

In terms of the implied correlation structure, as displayed in \autoref{tab:rho}, similar observations can be made. Indeed, there is increased variability in estimating the pairwise correlations $\rho_{jk}$ based on the method of moments in comparison to the likelihood-based methods. The three likelihood-based methods yield similar results both in terms of the estimates of $\rho_{jk}$ as well as the variability in these estimates. Naturally, the method of moments leads to estimates of the implied pairwise correlations $\rho_{jk}$ that are closer to the empirical estimates $r_{jk}$ as this method relies on matching sample moments to the theoretical moments. 

Overall, the data application further confirms the properties of the model that were exhibited through the simulation studies. Moreover, the illustration considered here demonstrates that even in the case of a small sample, with $n=71$ here, all estimation methods perform well.

\begin{table}[!htbp] \centering 
  \caption{Estimated values of $\Omega$ using the four estimation methods. Bootstrap standard errors are provided in parentheses.} 
  \label{tab:omega} 
\begin{tabular}{@{\extracolsep{5pt}} ccccc} 
\\[-1.8ex]\hline 
\hline \\[-1.8ex] 
 parameter& MM & SQ & 2S & ML \\ 
\hline \\[-1.8ex] 
$\omega_{21}$ & $0.447$ ($0.178$)& $0.332$ ($0.074$)& $0.330$ ($0.074$)& $0.310$ ($0.074$)\\ 
$\omega_{31}$ & $0.356$ ($0.174$)& $0.294$ ($0.085$)& $0.294$ ($0.083$)& $0.295$ ($0.085$)\\ 
$\omega_{32}$ & $0.030$ ($0.047$) & $0.095$ ($0.049$)& $0.076$ ($0.050$)& $0.083$ ($0.049$)\\ 
\hline \\[-1.8ex] 
\end{tabular} 
\end{table} 

\begin{table}[!htbp] \centering 
  \caption{Estimated values of $\Lambda$ for the two estimation methods. Bootstrap standard errors are provided in parentheses.} 
  \label{tab:lambda} 
\begin{tabular}{@{\extracolsep{5pt}} ccc} 
\\[-1.8ex]\hline 
\hline \\[-1.8ex] 
 parameter& MM & ML \\ 
\hline \\[-1.8ex] 
$\lambda_1$ & $7.366$ ($0.323$) & $7.353$ ($0.323$) \\ 
$\lambda_2$ & $13.408$ ($0.431$) & $13.357$ ($0.431$) \\ 
$\lambda_3$ & $10.662$ ($0.373$) & $10.611$ ($0.376$) \\ 
\hline \\[-1.8ex] 
\end{tabular} 
\end{table} 

\begin{table}[!htbp] \centering 
  \caption{Pairwise correlations based on the estimated model parameters with corresponding 95\% bootstrap percentile confidence intervals using the four estimation methods. } 
  \label{tab:rho} 
\begin{tabular}{@{\extracolsep{5pt}} cccccc} 
\\[-1.8ex]\hline 
\hline \\[-1.8ex] 
$\rho_{jk}$  & $r_{jk}$ &MM & SQ & 2S & ML \\ 
\hline \\[-1.8ex] 
$\rho_{12}$ & $0.625$ & $0.659$ & $0.568$ & $0.566$ & $0.548$ \\ 
& & $(0.406,0.933)$&  $(0.436,0.699)$&  $(0.435,0.697)$&  $(0.436,0.699)$\\ 
$\rho_{13}$ &$0.570$ & $0.586$ & $0.532$ & $0.532$ & $0.532$ \\ 
 & & $(0.330,0.891)$&  $(0.422,0.703)$&  $(0.427,0.703)$&  $(0.424,0.704)$\\ 
$\rho_{23}$ &$0.494$ & $0.494$ & $0.539$ & $0.512$ & $0.517$ \\ 
& & $(0.269,0.823)$&  $(0.385,0.690)$& $(0.364,0.697)$&  $(0.371,0.690)$\\ 
\hline \\[-1.8ex] 
\end{tabular} 
\end{table}

\section{Conclusion and discussion}\label{Sec:conclusion}

In this work, a new multivariate Poisson model is developed based on convolutions of multivariate comonotonic shocks. The proposed construction yields positively correlated Poisson random variables with a flexible dependence structure. Indeed, it is shown that the pairwise associations can freely range from independence to comonotonicity, thereby allowing for considerable flexibility in representing various strengths of dependence. Four estimation methods were explored in the proposed $\mathcal{MP}_d(\Lambda,\Omega)$ model, all of which were shown to perform well through several simulations as well as a real data analysis. 

While the proposed model contributes to the development of flexible multivariate Poisson models, there are, of course, further improvements that can be made and avenues for future work. For example, the present paper did not formally explore a regression-based version of the model. Incorporating covariate effects in the $\mathcal{MP}_d(\Psi)$ model is quite straightforward within the likelihood-based estimation procedures. For example, one can consider a formulation wherein the marginal Poisson rates vary according to a set of covariates $\mathbf{W}$, such that $\lambda_j=\exp(\mathbf{W}^{\top} \beta_j)$, where $\beta_j$, $j=1,\ldots,d$, represent the margin-specific regression coefficients. All three likelihood-based methods can then be adapted by reparameterizing each $\lambda_j$ in terms of $\beta_j$, for $j=1,\ldots,d$. In particular, the two-step and sequential likelihood methods would be straightforward as each $\beta_j$ could be estimated via univariate Poisson generalized linear models. Other variations could also be considered, e.g., by allowing the dependence parameters to vary in terms of covariate effects. Of course, sufficient data is required in order for estimation to be feasible. 

An important consideration in high dimensional problems is the numerical complexities that ensue. The proposed $d$-variate Poisson model, as defined in Equation~\eqref{equ:model}, is based on the convolutions of a set of $d$ comonotonic shock vectors of varying dimensions, and involves a total $d+d(d-1)/2$ parameters. As $d$ increases, simplifications may be necessary in order to reduce the model complexity. Indeed, in higher dimensions, a simplified version of the model based on $k<d$ comonotonic shock vectors could be considered, where $k$ is chosen so as to adequately capture the underlying dependence structure. In the sequential estimation procedures, such as the method of moments and the composite-likelihood approach, $k$ could be selected such that the remaining weight parameters $\omega_{kj}$, $j<k \in \{1,\ldots,d\}$ are below a small threshold $\varepsilon \in (0,1)$, where $\varepsilon$ could be chosen in a data-driven manner. 

A potential issue in using the proposed modelling framework in real data applications is the lack of robustness to outliers. It was found through numerical explorations that outlying observations can sometimes have a strong impact on estimation. This issue was particularly problematic in likelihood-based estimation due to the joint behavior induced by comonotonicity. Identifying and accounting for such outliers in the context of multi-dimensional discrete data is not immediately obvious and an area of future work. One possible solution is to consider a contamination type model, along the lines of \cite{Huber:1964}, wherein a mixture of the proposed $\mathcal{MP}_d(\Lambda,\Omega)$ model and an independence model could be used.

The lack of robustness to outlying observations can also lead to difficulties in establishing an appropriate ordering in the proposed model. 
%The lack of robustness to outlying \hl{or extremal} observations also contributes to an ordering problem in the proposed model. 
Although it is clear that permuting the margins changes the underlying model, such re-orderings of the components may sometimes distort the implied correlation structure. This problem did not appear in the data analysis, that is, similar estimates for the pairwise correlations ensued regardless of the ordering considered. Nonetheless, further numerical explorations were considered to study the impact of permutations in the proposed model. 
In considering a preliminary estimation of the first set of weights $\{\omega_{j1},j=1,\dots,d\}$ for all possible orderings, it was found that the number of potential pairwise outliers could be reduced through certain permutations of the margins. This would suggest an optimal ordering such that the impact of outlying observations is either eliminated, or minimized. In situations where such permutation issues arise, method of moments estimation may be preferable.
An alternative data-driven approach is to consider an ordering which aligns with the strength of the pairwise correlations. That is, the first component, $X_1$, could be chosen as that with the largest set of pairwise correlations, followed by the second component with the next largest set of pairwise correlations, and so on. Such an approach is particularly intuitive in the context of sequential estimation procedures (MM and SQ) as this ordering would ensure that the bulk of the pairwise associations are captured in earlier estimation steps of the elements of $\Omega$. This idea also aligns with our recommendation for a simplified $d$-dimensional model based on $k<d$ comonotonic shock vectors.

Finally, in its present form, the proposed model construction only allows for positive dependence. While negative dependence can be induced by incorporating counter-monotonic shocks into the construction, as is done in the bivariate setting in \cite{Genest/Mesfioui/Schulz:2018}, this idea does not easily extend to higher dimensions. Indeed, the notion of counter-monotonicity for dimension $d>2$ is not as straightforward. Some recent work has been done on this front, see, e.g., \cite{Lauzier:2023}. We leave further investigations of the case of negative dependence for future work.

\paragraph*{Data Availability}
The rainfall data explored in \autoref{Sec:data} can be downloaded from the Environment Canada weather archives \url{https://climate.weather.gc.ca/historical_data/search_historic_data_e.html}. The data compiled historic records from three weather stations in Nova Scotia (date accessed: September 22, 2023), namely Annapolis Royal (Climate ID 8200100), Springfield (Climate ID 8205200), and Kentville (Climate ID 8202810 and 8202800). 

  %% FUNDING INFORMATION (optional - comment out if unused). Provide
  %% any funding information as free form text. Do not abbreviate the
  %% names of granting agencies. You may specify a different title for
  %% this information through the optional argument.
\paragraph*{Funding}
  This work was supported by the Natural Sciences and Engineering Research Council of Canada Discovery Grants.
\paragraph*{Declarations}
The authors have no relevant financial or non-financial interests to disclose.

%% Appendices (optional - comment out if unused)
\appendix                       % start of appendices

\section{Proof of \autoref{prop_pqd}}\label{Proof}

The following provides a proof of the pairwise PQD ordering property established in \autoref{prop_pqd}.

\begin{proof}

Begin by writing 
\[
(X_i,X_j)=(Y_i+Z_{ik}+W_i,Y_j+Z_{jk}+W_j)
\]
where 
\begin{itemize}
\item $(Y_i,Y_j) = \left(\sum_{\ell=1,\ell\neq k}^i Z_{i \ell}, \sum_{\ell=1,\ell\neq k}^i Z_{j \ell} \right)$,
\item $(Z_{i\ell},Z_{j\ell})\sim \mathcal{M}\left\{\mathcal{P}(\omega_{i\ell}\lambda_i),\mathcal{P}(\omega_{j\ell}\lambda_j) \right\}$ for $\ell \in \{1,\ldots,i\}$,
\item $(Z_{i r},Z_{jr}) \perp (Z_{i s},Z_{j s})$ for any $r\neq s \in \{1,\ldots,i \}$,
\item $W_i \sim \mathcal{P}(0)$, i.e., a degenerate random variable with mass at $0$, which is independent of all $Z_{r\ell}$, $\ell \in\{1,\ldots,i\}$, $r=i,j$,
\item $W_j \sim \mathcal{P}\{(1-\sum_{\ell=1}^i \omega_{j\ell})\lambda_j \}$, which is independent of all $Z_{r\ell}$, $\ell \in\{1,\ldots,i\}$, $r=i,j$.
\end{itemize}
Similarly, write $(X_i^\prime,X_j^\prime)=(Y_i^\prime+Z_{ik}^\prime+W_i^\prime,Y_j^\prime+Z_{jk}^\prime+W_j^\prime)$, using analogous definitions for the components $Y_r^\prime$, $Z_{rk}^\prime$, $W_r^\prime$, $r=i,j$, in terms of the same parameters $\Psi_{(i,j)}$, with the exception of the modified weight $\omega_{jk}^{\prime}$.

For each $\ell\in\{1,\ldots,i\}\setminus k$, $(Z_{i\ell},Z_{j\ell}) \overset{d}{=} (Z_{i\ell}^\prime,Z_{j\ell}^\prime)$ and thus, trivially, one can write $(Z_{i\ell},Z_{j\ell}) \prec_{PQD} (Z_{i\ell}^\prime,Z_{j\ell}^\prime)$. 
As shown in Theorem~9.A.1 of \cite{Shaked/Shanthikumar:2007}, the PQD ordering is closed under convolutions, and thus $(Y_i,Y_j) \prec_{PQD} (Y_i^\prime,Y_j^\prime)$. Note that both $Y_i$ and $Y_i^\prime$ are Poisson-distributed with rate $\sum_{\ell=1,\ell\neq k}^i \omega_{i \ell} \lambda_i$, while both $Y_j$ and $Y_j^\prime$ have distribution $\mathcal{P}(\sum_{\ell=1,\ell \neq k}^i \omega_{j\ell}^\prime \lambda_j)$.

Next, rewrite $W_j$ as
\[
W_j=T_j+S_{jk},
\]
where $T_j \sim \mathcal{P}\{(1-\sum_{\ell=1}^i \omega_{j\ell}^\prime)\lambda_j\}$ is independent of $S_{jk} \sim \mathcal{P}\left\{(\omega_{jk}^\prime-\omega_{jk})\lambda_j \right\}$. Now recall the comonotonic pairs $(Z_{ik},Z_{jk}) \sim \mathcal{M}\left\{ \mathcal{P}(\omega_{ik}\lambda_i),\mathcal{P}(\omega_{jk}\lambda_j) \right\}$ and $(Z_{ik}^\prime,Z_{jk}^\prime) \sim \mathcal{M}\left\{ \mathcal{P}(\omega_{ik}\lambda_i),\mathcal{P}(\omega_{jk}^\prime\lambda_j) \right\}$. We then have that $(Z_{ik},Z_{jk}+S_{jk}) \prec_{PQD} (Z_{ik}^\prime,Z_{jk}^\prime)$ as both pairs retain the same marginal distributions, and $(Z_{ik}^\prime,Z_{jk}^\prime)$ is comonotonic, thereby reaching the Fréchet-Hoeffding bound given in \eqref{equ:FH_bounds}. We can also trivially ascertain that $(W_i,T_j)\prec_{PQD} (W_i^\prime,W_j^\prime)$ as both consist of independent components with the same margins, namely, $\mathcal{P}(0)$ and $\mathcal{P}\{(1-\sum_{\ell=1}^i \omega_{j\ell}^\prime )\lambda_j\}$, respectively.

Then, using once again the closure property of the PQD ordering (Theorem~9.A.1 of \cite{Shaked/Shanthikumar:2007}), we have that
\[
(Y_i+Z_{ik}+W_i,Y_j+Z_{jk}+T_j+S_{jk}) \prec_{PQD} (Y_i^\prime+Z_{ik}^\prime+W_i^\prime,Y_j^\prime+Z_{jk}^\prime+W_j^\prime) .
\]
This completes the proof as $(Y_i+Z_{ik}+W_i,Y_j+Z_{jk}+T_j+S_{jk})=(X_i,X_j)$ and $(Y_i^\prime+Z_{ik}^\prime+W_i^\prime,Y_j^\prime+Z_{jk}^\prime+W_j^\prime)=(X_i^\prime,X_j^\prime)$.

\end{proof}

\section{Asymptotic properties of estimators}\label{Appendix}

The following work provides further details on the asymptotic properties of the method of moments estimators (\autoref{Sec:MM}), two-step likelihood estimators (\autoref{Sec:2S}) and the sequential likelihood estimators (\autoref{Sec:SQ}). 

\subsection{Method of moments estimators}\label{App:MM}

As detailed in \autoref{Sec:MM}, method of moments estimation allows for a simplified implementation wherein each component of the parameter vector $\Psi$ is estimated sequentially. Despite the sequential implementation, the MM estimators $\mathring{\Psi}$ are in fact the simultaneous solutions to a set of estimating equations. Define $\mathbf{M}(\mathbf{X};\Psi)$ as 
\[
\mathbf{M}(\mathbf{X};\Psi)=
\begin{bmatrix}
X_1-\lambda_1 \\
\vdots \\
X_d - \lambda_d \\
(X_1-\lambda_1)(X_2-\lambda_2)-\sigma_{12}\\
\vdots \\
(X_{d-1}-\lambda_{d-1})(X_{d}-\lambda_d)- \sigma_{d-1 d}
\end{bmatrix}
\]
where, for ease of notation, $\sigma_{ij}$ is used to denote the covariance of the pair $(X_i,X_j)$, i.e.,
\[
\sigma_{ij}=\Cov(X_i,X_j)=\sum_{k=1}^{\min(i,j)} m_{\lambda_i,\lambda_j}(\omega_{ik},\omega_{jk}).
\]
It is clear that $\E\left\{ \mathbf{M}(\mathbf{X};\Psi_0) \right\}=\mathbf{0}$ where $\Psi_0$ denotes the true value of the model parameters. Moreover, as outlined in \autoref{Sec:MM}, solving the estimating equation $\sum_{i=1}^n \mathbf{M}(\mathbf{X}_i;\Psi)=\mathbf{0}$ leads to a unique solution corresponding to the MM estimators $\mathring{\Psi}$. It then follows from the theory of M-estimation, or estimating equations (see, e.g., \cite{Essential:2013}), that $\mathring{\Psi} \overset{P}{\rightarrow} \Psi_0$ and $\sqrt{n}(\mathring{\Psi}-\Psi_0) \rightsquigarrow \mathcal{N}_D \{\mathbf{0},V_M(\Psi_0)\}$ where the asymptotic variance is given by $V_M(\Psi_0)=A_M(\Psi_0)^{-1} B_M(\Psi_0) \{A_M(\Psi_0)^{-1}\}^{\top}$ and
\begin{align*}
A_M(\Psi_0) &= \E_{\Psi_0}\left\{ - \,\frac{\partial}{\partial\, \Psi^{\top}} \mathbf{M}(\mathbf{X};\Psi) \right\} \\
B_M(\Psi_0) &= \E_{\Psi_0}\left\{ \mathbf{M}(\mathbf{X};\Psi) \mathbf{M}(\mathbf{X};\Psi) ^{\top}  \right\} .
\end{align*}
While empirical estimates for $A_M(\Psi_0)$ and $B_M(\Psi_0)$ can be used, in general, this is difficult to evaluate. As such, standard bootstrap methods can be used to estimate $V_M(\Psi_0)$. 

Note that simplifications are possible for the diagonal components of $V_M(\Psi_0)$ in the proposed model. It is clear that for each $j=1,\ldots,d$, $\mathring{\lambda}_j=\bar{X}_j$ and thus $\sqrt{n}(\mathring{\lambda}_j-\lambda_{j0}) \rightsquigarrow \mathcal{N}(0,\lambda_{j0})$, where $\lambda_{j0}$ is used to denote the true value of the $j^{th}$ marginal parameter $\lambda_j$. As shown in, e.g., \cite{Ferguson:1996} (see Theorem~8 on p.~52), 
\[
\sqrt{n}(S_{jk}-\sigma_{jk0}) \rightsquigarrow \mathcal{N}(0,\Var\left\{ (X_j-\lambda_{j0}) (X_k-\lambda_{k0}) \right\})
\]
where $\sigma_{jk0}$ represents the true value of the pairwise covariance $\Cov(X_i,X_j)$. Recall that when estimating the weight parameters, at step $k$, $k=1,\ldots,d-1$, each element in $\Omega_k=\{\omega_{jk}, j=k+1,\ldots,d\}$ is individually estimated from distinct components of the set of estimating equations $\mathbf{M}(\mathbf{X};\Psi)$. In particular, $\omega_{jk}$ is estimated by the solution to $S_{jk}=\sigma_{jk}$, holding all other parameters fixed at the estimates obtained from previous steps in the estimation process. For ease of notation, write 
\[
M_{\Psi(j,k)}(\omega_{jk}) = \sum_{i=1}^{k} m_{\lambda_j,\lambda_k}(\omega_{ji},\omega_{ki})
\]
to emphasize that when estimating $\omega_{jk}$, all other parameters in $\Psi_{(j,k)}=(\lambda_j,\lambda_k,\omega_{k \ell}, \omega_{j \ell}, \ell=1,\ldots,k)$ are held fixed at their respective MM estimate, for each $k<j \in \{1,\ldots,d\}$. Since $S_{jk}$ is asymptotically normal, applying the Delta Method allows to establish the asymptotic normality of the MM estimator $\mathring{\omega}_{jk}$. Indeed, letting $\gamma_{jk}: \omega_{jk} \mapsto M_{\Psi(j,k)}^{-1}(\omega_{jk})$, one has that 
\[
\sqrt{n}(\mathring{\omega}_{jk}-\omega_{jk0}) \rightsquigarrow \mathcal{N} \left[ 0,\{\gamma_{jk}^{\prime}(\sigma_{jk0})\}^2 \Var\left\{(X_j-\lambda_j)(X_k-\lambda_k)\right\} \right]
\]
where $\gamma_{jk}^{\prime}$ is used to denote the derivative of $\gamma_{jk}$. Evaluating the above asymptotic variance is, in general, difficult and so standard bootstrap techniques are recommended.

\subsection{Two-step likelihood-based estimation}\label{App:2S}

The IFM estimators discussed in \autoref{Sec:2S} involve a two-step procedure wherein the marginal parameters are first estimated via their marginal MLEs and in the next step the dependence parameters are estimated by maximizing the full model log-likelihood. Define $\mathbf{Q}(\mathbf{X};\Psi)$ as
\[
\mathbf{Q}(\mathbf{X};\Psi) = 
\begin{bmatrix}
\frac{\partial}{\partial \, \Lambda^{\top}} \sum_{j=1}^d \log g_{\lambda_j}(X_j) \\[1em]
\frac{\partial}{\partial \, \Omega^{\top}} \log f_{\Psi}(\mathbf{X})
\end{bmatrix} .
\]
Under the usual regularity conditions, the IFM estimators $\check{\Psi}$ are in fact the solution to $\sum_{i=1}^n \mathbf{Q}(\mathbf{X}_i;\Psi)=\mathbf{0}$. Once again, the theory of M-estimation ensures that $\check{\Psi}$ is consistent and asymptotically Gaussian with 
\[
\sqrt{n}(\check{\Psi}-\Psi_0) \rightsquigarrow \mathcal{N}_D\left\{\mathbf{0},V_Q(\Psi_0)\right\}
\]
where $V_Q(\Psi_0)$ has the usual sandwich form stemming from the underlying estimating equation $\mathbf{Q}(\mathbf{X};\Psi)$, i.e., $V_Q(\Psi_0)=A_Q(\Psi_0)^{-1} B_Q(\Psi_0) \{A_Q(\Psi_0)^{-1}\}^{\top}$ with
\begin{align*}
A_Q(\Psi_0) &= \E_{\Psi_0}\left\{ - \,\frac{\partial}{\partial\, \Psi^{\top}} \mathbf{Q}(\mathbf{X};\Psi) \right\} \\
B_Q(\Psi_0) &= \E_{\Psi_0}\left\{ \mathbf{Q}(\mathbf{X};\Psi) \mathbf{Q}(\mathbf{X};\Psi) ^{\top}  \right\} .
\end{align*}

As shown in \cite{Joe:2005}, $V_Q(\Psi_0)$ can be further simplified here as $\mathbf{Q}(\mathbf{X};\Psi)$ is comprised of the marginal and joint model score equations. To see this, first, consider a decomposition of the full model information matrix as follows

\[
\mathcal{I} = 
\begin{bmatrix}
\mathcal{I}_{11} & \cdots & \mathcal{I}_{1d} & \mathcal{I}_{1w} \\
\vdots & \ddots & \vdots & \vdots \\
\mathcal{I}_{d1} & \cdots & \mathcal{I}_{dd} & \mathcal{I}_{dw} \\
\mathcal{I}_{w1} & \cdots & \mathcal{I}_{wd} & \mathcal{I}_{ww} 
\end{bmatrix}
\]
where
\begin{align*}
\mathcal{I}_{jk} &= -\E \left\{ \partial^2 \log f_{\Psi}(\mathbf{X})  / \partial \, \lambda_j \partial \, \lambda_k \right\}, \quad j,k \in \{1,\ldots,d\}, \\
\mathcal{I}_{jw} &= -\E \left\{ \partial^2 \log f_{\Psi}(\mathbf{X}) / \partial \, \lambda_j \partial \, \Omega^{\top} \right\}, \quad j \in \{1,\ldots,d\}, \\
\mathcal{I}_{wj} &= \mathcal{I}_{jw}^{\top}, \\
\mathcal{I}_{ww} &= -\E \left\{ \partial^2 \log f_{\Psi}(\mathbf{X}) / \partial \, \Omega \partial \, \Omega^{\top} \right\} .
\end{align*}

Write the underlying score equations as 
\[
\mathbf{Q}(\mathbf{X};\Psi)^{\top}=\left[ Q_1(X_1;\lambda_1),\ldots, Q_d(X_d;\lambda_d), Q_w(\mathbf{X};\Psi) \right]
\]
such that $Q_j(X_j;\lambda_j)=\partial \log f_j(X_j) / \partial \lambda_j \, $ consists of the marginal score equations, $j=1,\ldots,d$, and $Q_w(\mathbf{X};\Psi)=\partial  \log f_{\Psi}(\mathbf{X}) / \partial \, \Omega^{\top} $. Define $\mathcal{J}_{jk} = \Cov\left\{ Q_j(X_j;\lambda_j),Q_k(X_k,\lambda_k) \right\} $, for each $j,k \in \{1,\ldots,d\}$. Note that $\mathcal{J}_{jj}$ consists of the information matrix stemming from the $j^{th}$ margin. With this notation, \cite{Joe:2005} shows that

\[
A_Q(\Psi_0) = 
\begin{bmatrix}
\mathcal{J}_{11} & \cdots & 0 & \mathbf{0} \\
\vdots & \ddots & \vdots & \vdots & \\
0 & \cdots & \mathcal{J}_{dd} & \mathbf{0} \\
\mathcal{I}_{w1} & \cdots & \mathcal{I}_{wd} & \mathcal{I}_{ww} 
\end{bmatrix}
, \;
A_Q(\Psi_0)^{-1} = 
\begin{bmatrix}
\mathcal{J}_{11}^{-1} & \cdots & 0 & \mathbf{0} \\
\vdots & \ddots & \vdots & \vdots & \\
0 & \cdots & \mathcal{J}_{dd}^{-1} & \mathbf{0} \\
a_1 & \cdots & a_d & \mathcal{I}_{ww}^{-1} 
\end{bmatrix}
\]
where $a_j = -\mathcal{I}_{ww}^{-1} \mathcal{I}_{wj} \mathcal{J}_{jj}^{-1}$ for $j=1,\ldots,d$, and 
\[
B_Q(\Psi_0) = \E_{\Psi_0}\left\{ \mathbf{Q}(\mathbf{X};\Psi) \mathbf{Q}(\mathbf{X};\Psi) ^{\top}  \right\} =
\begin{bmatrix}
\mathcal{J}_{11} & \cdots & \mathcal{J}_{1d} & \mathbf{0} \\
\vdots & \ddots & \vdots & \vdots \\
\mathcal{J}_{d1} & \cdots & \mathcal{J}_{dd} & \mathbf{0} \\
\mathbf{0} & \cdots & \mathbf{0} & \mathcal{I}_{ww}
\end{bmatrix} .
\]

In the proposed model, for each $j,k \in \{1,\ldots,d\}$
\begin{align*}
\mathcal{J}_{jj} &= 1/\lambda_j \\
a_j & = -\lambda_j \mathcal{I}^{-1}_{ww}\mathcal{I}_{wj} \\
\mathcal{J}_{jk} &= M(\Psi_{(j,k)})/ \lambda_j \lambda_k
\end{align*}
where the latter uses $M(\Psi_{(j,k)})$ to denote $\Cov(X_j,X_k)$. From this, the asymptotic variance  $V_Q(\Psi_0)$ can be further simplified, as shown in \cite{Joe:2005}. Let $V_{Q,jk}$ denote the $(j,k)^{th}$ entry of $V_Q(\Psi_0)$. For $j,k \in \{1,\ldots,d\}$, the proposed model then leads to
\begin{align*}
V_{Q,jj} &= \lambda_j \\
V_{Q,jk} &= M(\Psi_{(j,k)}) \\
V_{Q,jw} &= -\sum_{k=1}^d M(\Psi_{(j,k)}) \mathcal{I}_{wk}^{\top} (\mathcal{I}^{-1}_{ww})^{\top} \\
V_{Q,wj} &= -\sum_{k=1}^d M(\Psi_{(j,k)}) \mathcal{I}_{ww}^{-1} \mathcal{I}_{wk} \\
V_{Q,ww} &= \mathcal{I}_{ww}^{-1} + \sum_{j=1}^d \sum_{k=1}^d -M(\Psi_{(j,k)}) \mathcal{I}_{ww}^{-1} \mathcal{I}_{wj} \mathcal{I}_{wk}^{\top} (\mathcal{I}_{ww}^{-1})^{\top} .
\end{align*}
As previously noted, in practice, bootstrap techniques can be used to approximate the above expressions.

\subsection{Sequential likelihood-based estimation}\label{App:SQ}

Recall from \autoref{Sec:SQ} that the sequential likelihood estimators $\tilde{\Psi}$ are the solutions to the set of estimating equations defined as
\[
\mathbf{G}(\mathbf{X};\Psi)=
\begin{bmatrix}
\mathbf{G}_{0}(\mathbf{X};\Lambda) \\
\mathbf{G}_1(\mathbf{X};\Omega_1) \\
\mathbf{G}_2(\mathbf{X};\Omega_2) \\
\vdots \\
\mathbf{G}_{d-1}(\mathbf{X};\Omega_{d-1}) 
\end{bmatrix}
=\mathbf{0},
\]
where 
\begin{align*}
\mathbf{G}_{0}(\mathbf{X};\Lambda) &= \left[ \frac{\partial}{\partial \, \Lambda^{\top}} \sum_{j=1}^d \log g_{\lambda_j}(X_j) \right] \\
&=\left[ \frac{\partial}{\partial \, \lambda_1} \log g_{\lambda_1}(X_1), \ldots, \frac{\partial}{\partial \, \lambda_d} \log g_{\lambda_d}(X_d) \right]^{\top}
\end{align*}
and for each $k=1,\ldots,d$
\begin{align*}
\mathbf{G}_k(\mathbf{X};\Omega_k) &=
\left[ \frac{\partial}{\partial \, \Omega_k^{\top}} \sum_{j=k+1}^d \log f_{k,j}(X_k,X_j) \right] \\
&= \left[ \frac{\partial}{\partial \, \omega_{k+1\, k}} \log f_{k,k+1}(X_k,X_{k+1}),\ldots,\frac{\partial}{\partial \, \omega_{d k}} \log f_{k,d}(X_k,X_d) \right]^{\top}.
\end{align*}
The sequential likelihood estimators $\tilde{\Psi}$ are such that $\sum_{i=1}^n \mathbf{G}(\mathbf{X}_i;\Psi)=\mathbf{0}$.

Under the usual regularity conditions in the context of likelihood-based estimation, specifically for the marginal and pairwise models, $\E\left\{\mathbf{G}(\mathbf{X};\Psi)\right\}=\mathbf{0}$. It then follows, once again, from the theory of M-estimation that 
\[
\sqrt{n}(\tilde{\Psi}-\Psi_0) \rightsquigarrow \mathcal{N}_D (\mathbf{0}, V_{G}(\Psi_0))
\]
where the asymptotic variance stems from the underlying estimating equations, viz. 
\[
V_G(\Psi_0)=A_G(\Psi_0)^{-1}B_G(\Psi_0)\{A_G(\Psi_0)^{-1}\}^{\top}
\]
with 
\begin{align*}
A_G(\Psi_0) &= \E_{\Psi_0}\left\{ - \,\frac{\partial}{\partial\, \Psi^{\top}} \mathbf{G}(\mathbf{X};\Psi) \right\} \\
B_G(\Psi_0) &= \E_{\Psi_0}\left\{ \mathbf{G}(\mathbf{X};\Psi) \mathbf{G}(\mathbf{X};\Psi) ^{\top}  \right\} .
\end{align*}

Based on the form of $\mathbf{G}(\mathbf{X};\Psi)$, $A_{G}(\Psi_0)$ can be partitioned as
\[
\begin{bmatrix}
\mathbf{A}_0 & \mathbf{0} \\
\mathbf{A}_{lw} & \mathbf{A}_{ww} .
\end{bmatrix}
\]
Similar to what was derived for the IFM estimators, $\mathbf{A}_0$ is a $d \times d$ matrix whose form stems from the likelihood under independence, yielding
\begin{align*}
\mathbf{A}_0 &= -\E_{\Psi_0} \left\{ \frac{\partial}{\partial\, \Lambda} \mathbf{G}_0(\mathbf{X};\Lambda) \right\} \\
&= -\E_{\Psi_0} \left\{ \frac{\partial^2}{\partial\, \Lambda \partial\, \Lambda^{\top}} \sum_{j=1}^d \log g_{\lambda_j}(X_j) \right\} \\
&= 
\begin{bmatrix}
\mathcal{J}_{11} &  \cdots & 0 \\
\vdots & \ddots & \vdots \\
0 & \cdots & \mathcal{J}_{dd}
\end{bmatrix} \\
&= 
\begin{bmatrix}
1/\lambda_{11} &  \cdots & 0 \\
\vdots & \ddots & \vdots \\
0 & \cdots & 1/\lambda_{dd}
\end{bmatrix} . 
\end{align*}
The block $\mathbf{A}_{ww}$ involves the dependence parameters $\Omega$ and consists of a $d(d-1)/2$ lower triangular matrix given by
\[
\mathbf{A}_{ww} = -\E_{\Psi_0} \left\{ \frac{\partial}{\partial \, \Omega} 
\left[ \mathbf{G}_1(\mathbf{X};\Omega_1),\ldots,G_{d-1}(\mathbf{X};\Omega_{d-1}) \right]^{\top}\right\}
\]
which simplifies to
\[
 -\E_{\Psi_0}
\begin{bmatrix}
\frac{\partial^2 \log f_{1,2}(X_1,X_2)}{\partial \omega_{21}^2}  & 0 & \cdots & 0 \\[1em]
\frac{\partial^2 \log f_{1,3}(X_1,X_3)}{\partial \omega_{21} \partial \omega_{31}}  & \frac{\partial^2 \log f_{1,3}(X_1,X_3)}{\partial \omega_{31}^2}   & \cdots & 0 \\[1em]
\vdots & \vdots & \ddots & \vdots \\[1em]
\frac{\partial^2 \log f_{d-1,d}(X_{d-1},X_d)}{\partial \omega_{21} \partial \omega_{d d-1}} & 
\frac{\partial^2 \log f_{d-1,d}(X_{d-1},X_d)}{\partial \omega_{31} \partial \omega_{d d-1}} & 
\cdots & 
\frac{\partial^2 \log f_{d-1,d}(X_{d-1},X_d)}{\partial \omega_{d d-1}^2}
\end{bmatrix}
\]
The block $\mathbf{A}_{lw}$ is a $d(d-1)/2 \times d$ matrix which can be further partitioned as
\[
\mathbf{A}_{lw}=
\begin{bmatrix}
\mathbf{A}_{lw,1} \\
\mathbf{A}_{lw,2} \\
\vdots \\
\mathbf{A}_{lw,d-1}
\end{bmatrix}
\]
where for each $k=1,\ldots,d-1$,
\begin{align*}
\mathbf{A}_{lw,k}&=-\E_{\Psi_0}\left\{ \frac{\partial }{\partial \Lambda}  \mathbf{G}_k(\mathbf{X};\Omega_k) \right\} \\
&= -\E_{\Psi_0} 
\begin{bmatrix}
\mathbf{A}_{lw,k}(0),\mathbf{A}_{lw,k}(1),\mathbf{A}_{lw,k}(2)
\end{bmatrix}
\end{align*}
and $\mathbf{A}_{lw,k}(0) = \mathbf{0}_{k-1}$ is a $d-k \times (k-1)$ matrix of $0$, $\mathbf{A}_{lw,k}(1)$ is a $d-k$ dimensional vector given by 
\[
\left[ \frac{\partial^2 \log f_{k,k+1}(X_k,X_k+1)}{\partial \lambda_k \partial \omega_{k+1 k}}, \ldots, \frac{\partial^2 \log f_{k,d}(X_k,X_d)}{\partial \lambda_k \partial \omega_{d k}} \right]^{\top},
\]
and $\mathbf{A}_{lw,k}(2)$ is a $(d-k)$ diagonal matrix equal to
\[
\text{diag}\left[ \frac{\partial^2 \log f_{k,k+1}(X_k,X_{k+1})}{\partial \lambda_{k+1} \partial \omega_{k+1 k}}, \ldots,\frac{\partial^2 \log f_{k,d}(X_k,X_{d})}{\partial \lambda_{d} \partial \omega_{d k}} \right] .
\]

The matrix $\mathbf{B}_G(\Psi_0)= \E_{\Psi_0}\left\{ \mathbf{G}(\mathbf{X};\Psi) \mathbf{G}(\mathbf{X};\Psi)^{\top} \right\} = \Cov_{\Psi_0}\left\{\mathbf{G}(\mathbf{X};\Psi)\right\} $ can be partitioned as follows
\[
\mathbf{B}_G(\Psi_0)
= 
\begin{bmatrix}
\mathbf{B}_0 & \mathbf{B}_{wl} \\
\mathbf{B}_{lw} & \mathbf{B}_{ww}
\end{bmatrix}.
\]
The upper left block $\mathbf{B}_0$ is a $d \times d$ matrix with $(j,k)^{th}$ element given by $\mathcal{J}_{jk}=M(\Psi_{(j,k)})/\lambda_j\lambda_k$ and diagonal elements $\mathcal{J}_{jj}=1/\lambda_j$, for $j,k=1,\ldots,d$, as defined for the IFM estimators in \autoref{App:2S}. $\mathbf{B}_{ww}$ is a $d(d-1)/2$ square matrix based on the score equations stemming from the bivariate likelihoods, i.e., 
\[
\mathbf{B}_{ww} = \Cov_{\Psi_0}\left\{ \left[\mathbf{G}_1(\mathbf{X};\Omega_1),\ldots,\mathbf{G}_{d-1}(\mathbf{X};\Omega_{d-1})  \right]^{\top} \right\}.
\]
The remaining blocks are given by 
\[
\mathbf{B}_{wl}=\mathbf{B}_{lw}^{\top} =\E_{\Psi_0} \left\{ \mathbf{G}_0(\mathbf{X};\Lambda) \left[\mathbf{G}_1(\mathbf{X};\Omega_1),\ldots,\mathbf{G}_{d-1}(\mathbf{X};\Omega_{d-1})  \right]^{\top} \right\}.
\]
Note that $\mathbf{B}_{lw} \neq \mathbf{0}$, in general, unlike the IFM estimators, see \autoref{App:2S}.

The asymptotic variance of the sequential likelihood estimators $\tilde{\Psi}$ then has the form
\begin{align*}
V_G(\Psi_0) &=A_G(\Psi_0)^{-1}B_G(\Psi_0)\{A_G(\Psi_0)^{-1}\}^{\top} \\
&= 
\begin{bmatrix}
\mathbf{A}_0^{-1} & \mathbf{0} \\
-\mathbf{A}_{ww}^{-1}\mathbf{A}_{lw}\mathbf{A}_0^{-1} & \mathbf{A}_{ww}^{-1} 
\end{bmatrix} 
\begin{bmatrix}
\mathbf{B}_0 & \mathbf{B}_{wl} \\
\mathbf{B}_{lw} & \mathbf{B}_{ww}
\end{bmatrix}
\begin{bmatrix}
(\mathbf{A}_0^{-1})^{\top} & (\mathbf{A}_0^{-1})^{\top}\mathbf{A}_{lw}^{\top}(-\mathbf{A}_{ww}^{-1})^{\top} \\
\mathbf{0} & (\mathbf{A}_{ww}^{-1})^{\top}
\end{bmatrix} \\
&= 
\begin{bmatrix}
\mathbf{V}_{G,11} & \mathbf{V}_{G,12} \\
\mathbf{V}_{G,21} & \mathbf{G}_{G,22} .
\end{bmatrix}
\end{align*}
In particular, $V_{G,11}$ is the asymptotic variance of the marginal parameter estimators, $\tilde{\Lambda}$, with 
\[
V_{G,11}=\mathbf{A}_0^{-1} \mathbf{B}_0(\mathbf{A}_0^{-1})^{\top}.
\]
Note that $V_{G,11}$ has the same form as the upper left $d \times d$ block of $V_Q$, that is, $V_{G,11}$ has $(j,k)^{th}$ elements given by $M(\Psi_{(j,k)})$ and diagonal elements $\lambda_j$, for $j,k=1,\ldots,d$; see \autoref{App:2S}. The block $V_{G,21}=V_{G,12}^{\top}$ simplifies to 
\[
-\mathbf{A}_{ww}^{-1} \left\{ \mathbf{A}_{lw}\mathbf{A}_0^{-1}\mathbf{B}_0-\mathbf{B}_{lw} \right\} (\mathbf{A}_0^{-1})^{\top}.
\]
Finally, $V_{G,22}$ is the asymptotic variance of the dependence parameters $\Omega$ with
\begin{small}
\[
\mathbf{A}_{ww}^{-1} \left\{ \mathbf{A}_{lw}\mathbf{A}_0^{-1}\mathbf{B}_0(\mathbf{A}_0^{-1})^{\top}\mathbf{A}_{lw}^{\top} - \mathbf{B}_{lw}(\mathbf{A}_0^{-1})^{\top} \mathbf{A}_{lw}^{\top} - \mathbf{A}_{lw}\mathbf{A}_0^{-1}\mathbf{B}_{wl} + \mathbf{B}_{ww} \right\} (\mathbf{A}_{ww}^{-1})^{\top}.
\]
\end{small}
As before, in practice, $V_G(\Psi_0)$ can be approximated using bootstrap techniques. 

\bibliographystyle{chicago}
\bibliography{MPoi_Ref}

\end{document}